\documentclass[a4,letters,usenatbib]{mnras}
\usepackage{amsmath}
\usepackage{amssymb} 
\usepackage{multirow}
\usepackage{textcomp}
\usepackage{microtype}
\usepackage{subcaption}
\usepackage{xcolor,ulem}
\usepackage{graphicx}
\usepackage{float}
\usepackage{mathptmx}

\newcommand{\email}[1]{\mbox{\href{mailto:#1}{#1}}}

\def\apj{Astrophysical Journal}                
\def\apjl{Astrophysical Journal}             

\def\apjs{ApJS}              
\def\mnras{MNRAS}            
\def\prd{Phys.~Rev.~D}       
\def\araa{ARA\&A}             
\def\nat{Nature}              
\def\aap{A\&A}                

\newcommand{\msun}{{\rm M}_\odot}

\title[GW background from eccentric MBHBs]{Efficient computation of the gravitational wave spectrum emitted by eccentric massive black hole binaries in stellar environments}
\author[S. Chen, A. Sesana, W. Del Pozzo]{
Siyuan Chen,$^1$\thanks{E-mail: \email{schen@star.sr.bham.ac.uk}}
Alberto Sesana$^1$\thanks{E-mail: \email{asesana@star.sr.bham.ac.uk}}
 and Walter Del Pozzo$^{1,2}$
\\
  $^1$School of Physics \& Astronomy, University of Birmingham, Birmingham, B15 2TT, UK\\
  $^2$Dipartimento di Fisica ``Enrico Fermi'', Universit\`a di Pisa, Pisa I-56127, Italy} 
\date{Accepted XXX. Received YYY; in original form ZZZ}

\pubyear{2017}

\begin{document}
\label{firstpage}
\pagerange{\pageref{firstpage}--\pageref{lastpage}}
\maketitle

\begin{abstract}

  We present a fast and versatile method to calculate the characteristic spectrum $h_c$ of the gravitational wave background (GWB) emitted by a population of eccentric massive black hole binaries (MBHBs). We fit the spectrum of a reference MBHB with a simple analytic function and show that the spectrum of any other MBHB can be derived from this reference spectrum via simple scalings of mass, redshift and frequency. We then apply our calculation to a realistic population of MBHBs evolving via 3-body scattering of stars in galactic nuclei. We demonstrate that our analytic prescription satisfactorily describes the signal in the frequency band relevant to pulsar timing array (PTA) observations. Finally we model the high frequency steepening of the GWB to provide a complete description of the features characterizing the spectrum. For typical stellar distributions observed in massive galaxies, our calculation shows that 3-body scattering alone is unlikely to affect the GWB in the PTA band and a low frequency turnover in the spectrum is caused primarily by high eccentricities.


\end{abstract}

\begin{keywords}
black hole physics -- gravitational waves -- galaxies: kinematics and dynamics -- methods: analytical
\end{keywords}

\section{Introduction}
\label{sec:Introduction}

It is now well established that most (potentially all) massive galaxies harbour massive black holes (MBHs) in their centre \citep[see][and references therein]{KormendyHo:2013}. In the standard hierarchical cosmological model \citep{WhiteRees:1978}, present day galaxies grow in mass and size by accreating cold gas from the cosmic web \citep{2009Natur.457..451D} and by merging with other galaxies \citep{1993MNRAS.264..201K}.  In a favoured scenario in which MBHs are ubiquitous up to high redshift, following the merger of two galaxies, the central MBHs hosted in their nuclei sink to the centre of the merger remnant eventually forming a bound binary system \citep{BegelmanBlandfordRees:1980}. The binary orbit shrinks because of energy and angular momentum exchange with the surrounding ambient, stars and cold gas \citep[see][for a recent review]{DottiSesanaDecarli:2012}, to the point at which gravitational wave (GW) emission takes over, efficiently bringing the pair to coalescence. Since galaxies are observed to merge quite frequently \citep[see, e.g.,][]{2006ApJ...652..270B,2009A&A...498..379D} and the observable Universe encompasses several billions of them, a sizeable cosmological population of MBHBs is expected to be emitting GWs at any time \citep[][hereinafter SVC08]{SesanaVecchioColacino:2008}.

At nHz frequencies, their signal is going to be captured by pulsar timing arrays \citep[PTAs][]{FosterBacker:1990}. Passing GWs leaves an imprint in the time of arrival of ultra-stable millisecond pulsars. By cross correlating data from an ensemble of millisecond pulsars (i.e. from a PTA), this signature can be confidently identified \citep{HellingsDowns:1983}. Because pulsars are timed on weekly basis ($\Delta{t}=1$week) over a period ($T$) of many years (almost 30yr for some of them), PTAs are sensitive to GW in the frequency window $[1/T,1/(2\Delta{t})]\approx [1{\rm nHz},1\mu{\rm Hz}]$.

The European Pulsar Timing Array \citep[EPTA][]{2016MNRAS.458.3341D} and the Parkes Pulsar Timing Array \citep[PPTA][]{2016MNRAS.455.1751R} and North American Nanohertz Observatory for Gravitational Waves \citep[NANOGrav][]{2015ApJ...813...65T} have made considerable advances in increasing the sensitivity of their datasets. And the first data release of the International Pulsar Timing Array \citep[IPTA][]{VerbiestEtAl_IPTA1stData:2016} is paving the way towards an effective combination of all PTA observation into a dataset that has the potential to detect GWs within the next ten years \citep{2015MNRAS.451.2417R,2016ApJ...819L...6T}. Moreover, new powerful telescopes such as the SKA pathfinder MeerKAT in South Africa \citep{2009arXiv0910.2935B} and the 500-mt FAST in China \citep{2011IJMPD..20..989N} will be online in the next couple of years, boosting the odds of GW detection with their exquisite timing capabilities.


The frequency spectrum of the overall GW signal is given by the superposition of all sources emitting at a given frequency. Because of the abundance of MBHBs in the Universe this has been generally described as a stochastic GW background (GWB), characterized, in case of circular GW driven binaries by a power-law spectrum $h_c\propto f^{-2/3}$ \citep{2001astro.ph..8028P}. However, two facts have became clear in the past decade. Firstly, to get to the PTA band, MBHBs need to efficiently interact with their stellar environment, which potentially has a double effect on the shape of the GW spectrum. If at the lower end of the PTA sensitivity window, MBHBs shrink because of interaction with their environment more efficiently than because of GW emission, then the spectrum is attenuated, even showing a potential turnover \citep{KocsisSesana:2011,Sesana:2013CQG,2014MNRAS.442...56R,2016arXiv160601900K}. Moreover, both scattering of ambient stars and interaction with a circumbinary disk tend to excite the binary eccentricity \citep{1996NewA....1...35Q,2005ApJ...634..921A,2011MNRAS.415.3033R}. This also results in a loss of power at low frequency (eccentric binaries evolve faster and generally emit at higher frequencies), potentially leading to a turnover in the spectrum \citep{EnokiNagashima:2007,HuertaEtAl:2015}. Secondly, for $f>10$nHz, the bulk of the signal is provided by sparse massive and/or nearby sources, and cannot be described simply as a stochastic signal. This was first noticed by SVC08, who came to the conclusion that at high frequency the GW signal will be dominated by sparse, individually resolvable sources, leaving behind a stochastic GWB at a level falling  below the nominal $f^{-2/3}$ power law.

With the constant improvement of their timing capabilities, PTAs are placing increasingly stringent limits on the amplitude of the expected GWB \citep{ShannonEtAl_PPTAgwbg:2015,LentatiEtAl_EPTAgwbg:2015,ArzoumanianEtAl_NANOGRAV9yrData:2016}, and detection is possible within the next decade. One crucial question is then: what astrophysics do we learn from a GWB detection with PTA? This question has been sparsely tackled by a number of authors \citep[see, e.g.,][]{Sesana:2013CQG} but the answers have been mostly qualitative. A full assessment of what we can learn from PTA detection will stem from a combination of all the measurements PTA will be able to make, including: amplitude and shape of the unresolved GW signal, possible non Gaussianity and non-stationarity, statistics and properties of individually resolvable sources. With this long-term goal in mind, a first step is to investigate what information can be retrieved from the {\it amplitude and shape} of the GWB.

As part of the common effort of the EPTA collaboration \citep{2016MNRAS.458.3341D} to detect GWs with pulsar timing, in this paper, we derive the expected spectrum of a GWB for a generic population of eccentric MBHBs evolving in typical stellar environments. Expanding on the work of \cite{MiddletonEtAl:2016}, the goal is to define a model that links the MBHB mass function and eccentricity distribution to the shape of the GWB spectrum. In particular, we find that the astrophysical properties of the MBHBs are reflected in two features of the spectrum. The efficiency of environmental coupling and the MBHB eccentricity might cause a low frequency flattening (or even a turnover) in the spectrum. The shape of the MBHB mass function affects the statistics of bright sources at high frequency, causing a steepening of the unresolved level of the GWB. We develop an efficient (mostly analytical) formalism to generate GW spectra given a minimal number of parameters defining the MBHB mass function, the efficiency of environmental coupling and the eccentricity distribution. In a companion paper we will show how the formalism developed here is suitable to an efficient exploration of the model parameter space, allowing, for the first time, a quantitative estimate of the MBHB population parameters from a putative GWB measurement. 

The paper is organized as follows. In Section \ref{sec:Model}, we derive a versatile and quick analytic approximation to the shape of a GW spectrum produced by eccentric GW driven binaries. In Section \ref{sec:Coupling} we study the evolution of eccentric MBHBs in stellar environments with properties constrained by observations of massive spheroids. We derive typical transition frequencies at which GWs take over, coalescence timescales, and we construct a simplified but robust framework to include the scattering-driven phase in the computation of the GW spectrum. Section \ref{sec:Population} reports the main results of our investigation. By employing a range of MBHB populations, we demonstrate that our quick approximation is applicable in the PTA frequency window, with little dependence on the detailed properties of the stellar environment. Moreover, we derive a fast way to compute the high frequency steepening of the spectrum to account for the small number statistics of massive, high frequency MBHBs. We discuss our results and describe future applications of our findings in Section \ref{sec:Conclusions}.

\section{Analytical modelling of the GW spectrum}
\label{sec:Model}

The GWB generated by a population of eccentric binaries was first investigated by \cite{EnokiNagashima:2007} and more recently by \cite{HuertaEtAl:2015}. In this section we follow the same approach and review their main results. Following \cite{2001astro.ph..8028P}, the characteristic strain $h_c(f)$ of the GW spectrum produced by a population of cosmological MBHBs  can be written as 

\begin{equation}
h_c^2(f) = \frac{4G}{\pi c^2 f} \int_{0}^{\infty} dz \int_{0}^{\infty} d{\cal M} \frac{d^2n}{dzd{\cal M}} \frac{dE}{df_r}. 
\label{eq:hc}
\end{equation}
Here, $d^2n/dzd{\cal M}$ defines the comoving differential number density (i.e. number of systems per Mpc$^3$) of merging MBHBs per unit redshift and unit chirp mass ${\cal M}=(M_1M_2)^{3/5}/(M_1+M_2)^{1/5}$ -- where $M_1>M_2$ are the masses of the binary components -- and the {\it observed} GW frequency at Earth $f$ is related to the {\it emitted} frequency in the source frame $f_r$ via  $f_r=(1+z)f$. The evaluation of equation (\ref{eq:hc}) involves a double integral in mass and redshift, generally to be performed numerically, and the computation of the energy spectrum $dE/df_r$. For an eccentric MBHB, this is given by a summation of harmonics as:
\begin{equation}
  \frac{dE}{df_r} = \sum_{n=1}^{\infty} \frac{1}{n} \frac{dE_n}{dt} \frac{dt}{de_n} \frac{de_n}{df_n},
  \label{eq:dEdf}
\end{equation}
where now $f_n=f_r/n$ is the restframe {\it orbital} frequency of the binary for which the $n$-th harmonic has an observed frequency equal to $f$ and $e_n$ is the eccentricity of the binary at that orbital frequency. We used the concatenation rule of derivation to highlight the role of the eccentricity. The first differential term in the rhs of equation (\ref{eq:dEdf}) is the luminosity of the $n$-th GW harmonic given by
\begin{equation}
  \frac{dE_n}{dt} = \frac{32}{5} \frac{G^{7/3}}{c^5} \mathcal{M}^{10/3} (2\pi f_n)^{10/3} g_n(e_n)
  \label{eq:dEdt}
\end{equation}
where
\begin{equation}
\begin{split}
& g_n(e) = \\ & \frac{n^4}{32} \Big[\Big(J_{n-2}(ne)-2eJ_{n-1}(ne)+\frac{2}{n}J_n(ne)+2eJ_{n+1}(ne)-J_{n+2}(ne)\Big)^2 \\ & +(1-e^2)\Big(J_{n-2}(ne)-2J_n(ne)+J_{n+2}(ne)\Big)^2 + \frac{4}{3n^2} J_n^2(ne)\Big],
\end{split}
\end{equation}
and $J_n$ is the $n$-th Bessel function of the first kind. The other two differential terms describe the evolution of the binary frequency and eccentricity with time, and for an eccentric MBHB driven by GW emission only, are given by
\begin{align}
\frac{df_n}{dt} & = \frac{96}{5} (2\pi)^{8/3} \frac{G^{5/3}}{c^5} \mathcal{M}^{5/3} f_n^{11/3} F(e_n) \label{eq:dfdt}
\\
\frac{de_n}{dt} & = -\frac{1}{15} (2\pi)^{8/3} \frac{G^{5/3}}{c^5} \mathcal{M}^{5/3} f_n^{8/3} G(e_n)
\label{eq:dedt}
\end{align}
where
\begin{align}
F(e) & = \frac{1+(73/24)e^2+(37/96)e^4}{(1-e^2)^{7/2}}
\\
G(e) & = \frac{304e+121e^3}{(1-e^2)^{5/2}}.
\end{align}
By plugging (\ref{eq:dEdt}), (\ref{eq:dfdt}) and (\ref{eq:dedt}) into the expression (\ref{eq:dEdf}), equation (\ref{eq:hc}) one obtains \citep{HuertaEtAl:2015}
\begin{equation}
\begin{split}
h_c^2(f) = & \frac{4G}{\pi c^2 f} \int_{0}^{\infty} dz \int_{0}^{\infty} d{\cal M}\frac{d^2n}{dzd{\cal M}} \\ & \frac{{\cal M}^{5/3}(\pi G)^{2/3}}{3(1+z)^{1/3} f^{1/3}}\sum_{n=1}^{\infty}\frac{g_n(e_n)}{F(e_n)(n/2)^{2/3}}.
\label{eq:hcgw}
\end{split}
\end{equation}

We note that Equation \eqref{eq:hcgw} is strictly valid only if the merger happens at a fixed redshift. However, we will see later that the typical merger timescale, $t_c$, of MBHBs can be Gyrs (cf Equation (\ref{eq:tcoal}) and figure \ref{sec:fttcsingle}), which is comparable to the cosmic expansion time $t_{\rm Hubble}$. Despite this fact, what actually matters is only the last phase of the MBHB inspiral, when the GW power is emitted in PTA band. Let us consider an optimistic PTA able to probe frequencies down to $\approx 1$nHz (i.e. observing for 30 years). If binaries are circular, then they start to emit in the PTA band only when their orbital frequency is $f_{\rm orb}=0.5$nHz. For typical MBHBs of ${\cal M}>3\times 10^{8}$ \citep[which are those dominating the GWB, see e.g.][]{SesanaVecchioColacino:2008}, the coalescence time at that point is $\tilde{t}_c<0.15$Gyr. The bulk of the PTA signal comes from $z<1.5$ \citep{2015MNRAS.447.2772R,2016ApJ...826...11S}, where the typical cosmic expansion time is already $t_{\rm Hubble}(z)>1$Gyr. This is almost an order of magnitude larger than $\tilde{t}_c$, which we also stress becomes much shorter with increasing MBHB masses. On the other hand, if binaries are very eccentric, they start to emit significant GW power in the PTA band when their orbital frequency is much lower than the minimum frequency probed by the array. Figure \ref{fig:speccompare} shows that, if $e=0.9$, considering only the power emitted since $f_{\rm orb}=0.1$nHz provides a good approximation to the overall spectrum from $f \approx 1$nHz onwards. Although $f_{\rm orb}$ is much lower in this case, eccentric binaries coalesce much faster (see again Equation (\ref{eq:tcoal})). For typical MBHBs of ${\cal M}>3\times 10^{8}$ with $e=0.9$, the coalescence time at that point is $\tilde{t}_c<10$Myr. Therefore, $\tilde{t}_c\ll t_{\rm Hubble}$ becomes a better approximation with increasing eccentricity, and Equation (\ref{eq:hcgw}) generally provide a good approximation to the GWB.
  


In practice, equation (\ref{eq:hcgw}) is evaluated numerically. For each integration element, the sum in the expression has to be computed by solving numerically equations (\ref{eq:dfdt}) and (\ref{eq:dedt}) to evaluate $e_n$ at each of the orbital frequencies $f_n$ contributing to the spectrum observed at frequency $f$, and by then computing the appropriate $g_n(e_n)$ function. This procedure is extremely cumbersome and time consuming. \citep{2009PhRvD..80h4001Y} proposed an analytical approximation for $e(f)$ that helps in speeding up the computation. However, it is accurate only for $e<0.9$, and even then one is left with the computation of the $n$ harmonics and the evaluation of the Bessel functions. Note that the GW energy spectrum of a binary with eccentricity $e$ peaks at the $n_p\approx(1-e)^{-3/2}$ harmonic, with still significant contributions at $n\sim 10n_p$ \citep{2010PhRvD..82j7501B}. For a MBHB with $e=0.9$ this implies the computation of several hundreds of harmonics.  

\subsection{Fitting formula and scaling properties}
\label{sec:fit}

\begin{figure}
\includegraphics[width=0.45\textwidth]{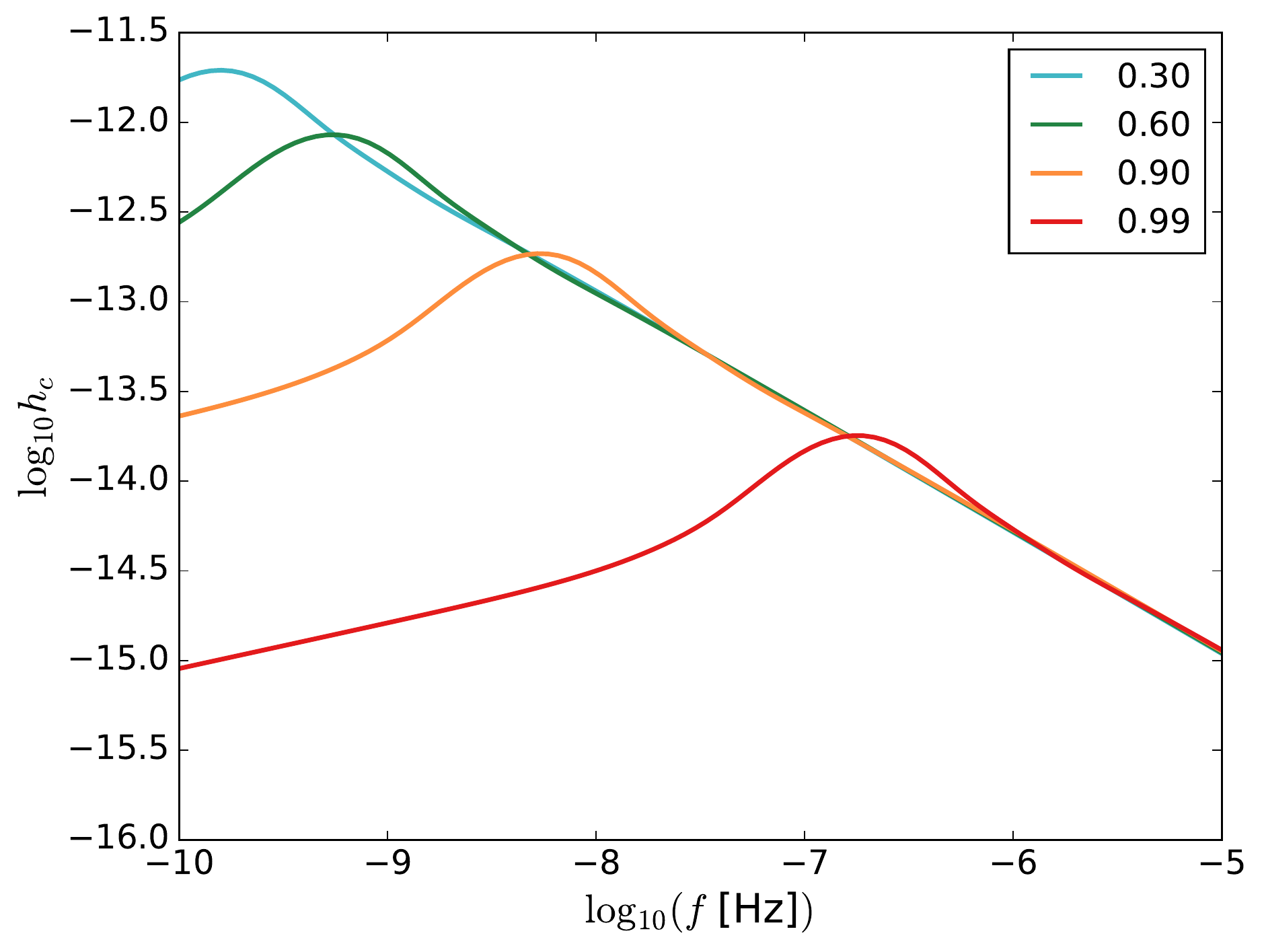}
\caption{characteristic amplitude spectrum for different eccentricities calculated with $n = 12500$ harmonics computed with no lower limit on $f_n$.}
  \label{fig:specint}
\end{figure}

Our first goal is to compute an efficient and accurate way to numerically calculate $h_c^2(f)$. Although the double integral might be solvable analytically for a suitable form of $d^2n/dzd{\cal M}$, a numerical evaluation is generally required. We therefore concentrate on the computation of the single integral element. We thus consider a reference system with a unity number density per Mpc$^3$ characterized by selected chirp mass and redshift. This corresponds to setting
\begin{equation}
\frac{d^2n}{dzd{\cal M}}=\delta({\cal M}-{\cal M}_0)\delta(z-z_0)/ \text{Mpc}^3.
\end{equation}
Equation (\ref{eq:hcgw}) then becomes
\begin{equation}
\begin{split}
  h_{c,0}^2(f) & = \frac{4G^{5/3} \text{Mpc}^{-3}}{3\pi^{1/3} c^2 f^{4/3}} \frac{{\cal M}_0^{5/3}}{(1+z_0)^{1/3}}\sum_{n=1}^{\infty}\frac{g(n,e_n)}{F(e_n)(n/2)^{2/3}}
\end{split}
  \label{eq:hc0}
\end{equation}
To fully define the system we need to specify an initial MBHB eccentricity $e_0$ at a reference orbital frequency $f_0$, so that the eccentricity $e_n=e_n(n,f_0,e_0)$ can to be evaluated for the appropriate $n-$th harmonic at the orbital frequency $f_n=f(1+z_0)/n$ via equations (\ref{eq:dfdt}) and (\ref{eq:dedt}).

\begin{figure}
\includegraphics[width=0.45\textwidth]{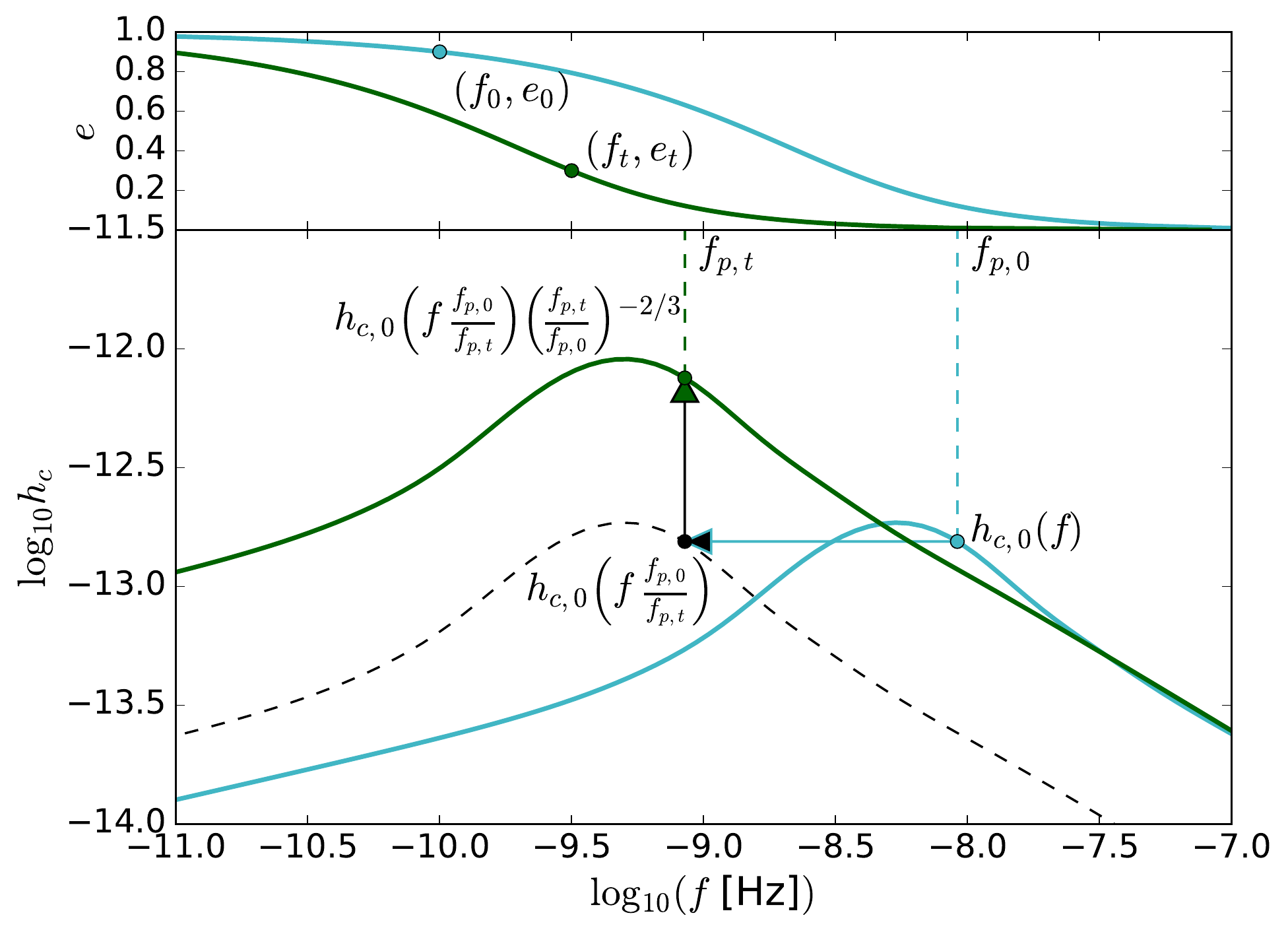}
\caption{Analytical spectral shift. The upper panel shows the eccentricity evolution over frequency for a fiducial spectrum characterized by the initial conditions $(e_0 = 0.9, \ f_0 = 10^{-10}$Hz$)$ (blue) and a generic spectrum characterized by $(e_t, \ f_t)$ (green). The lower panel shows the respective GW spectra (again, blue for fiducial and green for generic) and the steps involved in the shifting. The two vertical dashed lines mark the 'peak frequencies' defined in \protect\cite{HuertaEtAl:2015}, the horizontal arrow shifts the fiducial spectrum by $f_{p,t}/f_{p,0}$ (black dashed spectrum), and the vertical arrow moves it up by a factor $(f_{p,t}/f_{p,0})^{-2/3}$, as described in the main text.}
  \label{fig:specshift}
\end{figure}

We study the behaviour of equation (\ref{eq:hc0}) by taking a fiducial binary with ${\cal M}_0=4.16\times10^8\msun$, $z_0=0.02$, $f_0=0.1$nHz and different eccentricities $e_0=0.3, 0.6, 0.9, 0.99$. Results are shown in figure \ref{fig:specint}. Obviously, since the binary circularizes because of GW emission, at high frequency all the spectra eventually sit on the same power law. Moreover, the spectra look self-similar, as also noted by \cite{HuertaEtAl:2015}. This property allows the spectra to be shifted on the $f^{-2/3}$ diagonal, given an analytic fitting expression for one reference spectrum. Self similarity has to be expected because equations (\ref{eq:dfdt}) and (\ref{eq:dedt}) combine to give \citep{EnokiNagashima:2007}
\begin{equation}
\frac{f}{f_0}=\left(\frac{1-e_0^2}{1-e^2}\left(\frac{e}{e_0}\right)^{12/19}\left(\frac{1+\frac{121}{304}e^2}{1+\frac{121}{304}e_0^2}\right)^{870/2299}\right)^{-3/2}.
\end{equation}
This means that the eccentricity evolution is just a function of the frequency ratio $f/f_0$ and there is no intrinsic scale in the problem. Any inspiral will thus pass through any given eccentricity at some frequency during the process. A reference binary with $e_0=0.9$ at $f_0=10^{-10}$Hz is simply an earlier stage in the evolution of a binary with a smaller $e$ at a higher $f$, see figure \ref{fig:specshift}. Therefore, the spectrum of a binary with a different initial eccentricity $e_t$ specified at a different initial frequency $f_t$ can be simply obtained by shifting the spectrum of the reference binary. What one needs to know is by how much the spectrum has to be shifted along the $f^{-2/3}$ diagonal. To answer this question we identify a reference point of the spectrum. The obvious choice is the peak frequency defined by \cite{HuertaEtAl:2015}. They showed that the deviation of the spectrum of an eccentric binary, defined by fixing the eccentricity $e$ at a given orbital frequency $f$, with respect to its circular counterpart peaks at a frequency $f_p$ given by\footnote{Note that $f_p$ does not coincide with the peak of the characteristic amplitude, as also clear from figure \ref{fig:specshift}.}
\begin{equation}
  \frac{f_p}{f} =  \frac{1293}{181} \Big[\frac{e^{12/19}}{1-e^2}\big(1+\frac{121e^2}{304}\big)^{870/2299}\Big]^{3/2} 
  \label{eq:fpeak}
\end{equation}

Let us consider two spectra as shown in figure \ref{fig:specshift}. The first one is a reference spectrum $h_{c,0}(f)$ defined by $e_0=0.9$ at $f_0=10^{-10}$Hz, the second one is a generic spectrum $h_c(f)$ characterized by a generic value of $e_t$ at a transition frequency $f_t$ typically different from $f_0$. By feeding these input into \eqref{eq:fpeak} we directly get the two peak frequencies $f_{p,0}$ and $f_{p,t}$ respectively, marked in the lower panel of the figure. We want to compute $h(f)$ from $h_{c,0}(f)$. It is clear that the peak frequency at $f_{p,0}$ has to shift to  $f_{p,t}$, therefore $h_c(f)$ has to correspond to $h_{c,0}(f')$ where $f'=f(f_{p,0}/f_{p,t})$. However, this transformation just shifts the spectrum horizontally. To get to $h_{c,0}(f)$ we still need to multiply $h_{c,0}(f')$ by a factor $(f_{p,t}/f_{p,0})^{-2/3}$. The total shift has therefore the form
\begin{equation}
  h_c(f) = h_{c,0}\Big(f\frac{f_{p,0}}{f_{p,t}}\Big)\left(\frac{f_{p,t}}{f_{p,0}}\right)^{-2/3},
  \label{eq:hshift}
\end{equation}
In fact, it is easy to verify that by applying equation (\ref{eq:hshift}) to any of the spectra in figure \ref{fig:specint} all the other spectra are recovered.

\begin{figure}
\includegraphics[width=0.45\textwidth]{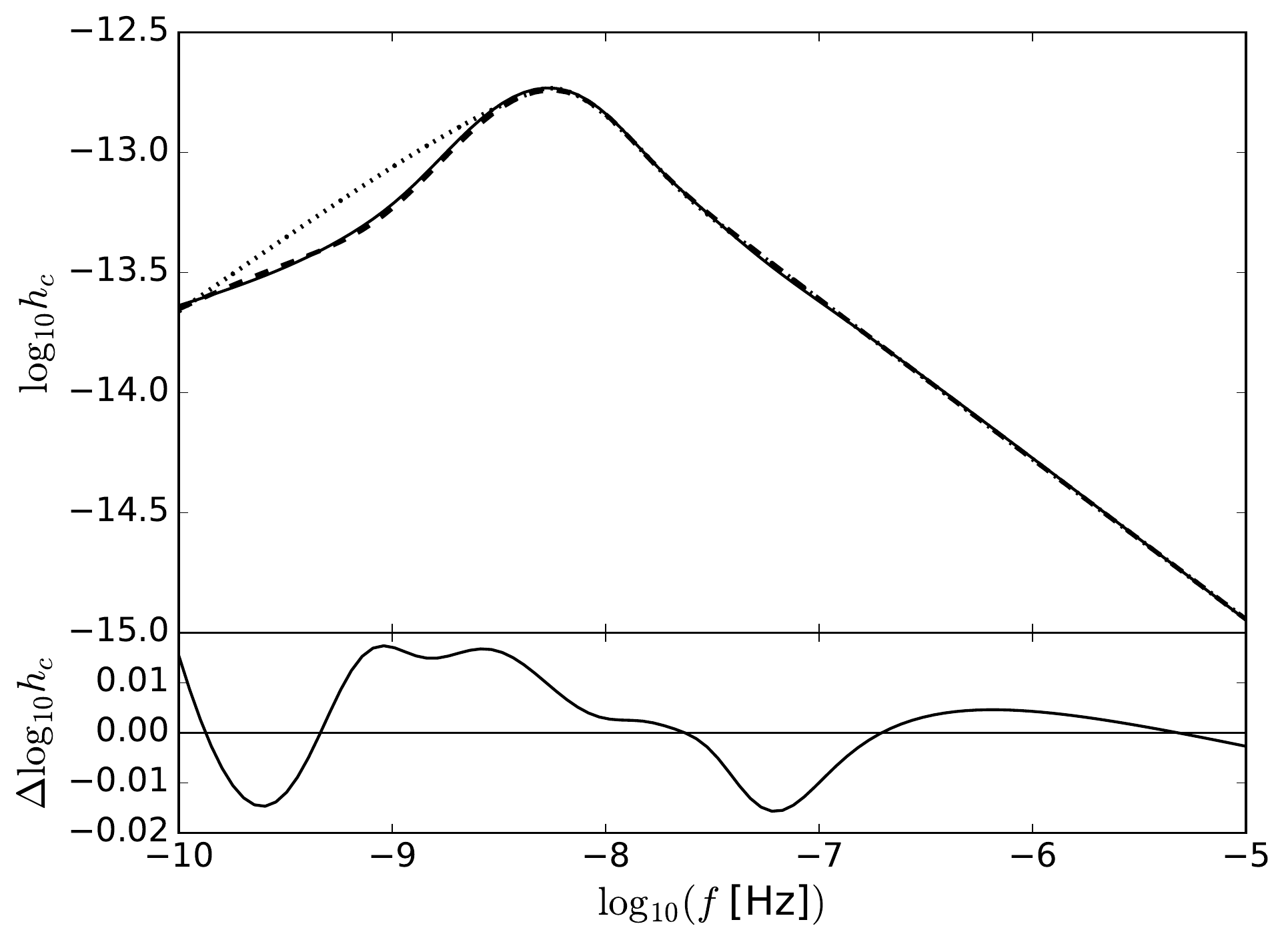}
\caption{Gravitational wave spectrum $h_c(f)$ for the reference binary described in the text, computed by summing $n = 12500$ harmonics (dashed line) compared to the best fit $h_{c,{\rm fit}}(f)$ with an analytic function of the form given by equation (\ref{eq:hcfit}) (solid line) and by \protect\cite{HuertaEtAl:2015} (dotted line). The lower panel shows the difference ${\rm log}_{10}h_{c,{\rm fit}}-{\rm log}_{10}h_{c}$ as a function of frequency.}
\label{fig:specfit}
\end{figure}

All we need then is a suitable fit for a reference MBHB. For this, we take the reference case $f_0=10^{-10}$Hz and $e_0=0.9$ and, based of the visual appearance on the spectrum, we fit a trial analytic function of the form
\begin{equation}
  h_{c,{\rm fit}}(f) = a_0 \bar{f}^{a_1} e^{-a_2 \bar{f}}+b_0 \bar{f}^{b_1} e^{-b_2 \bar{f}}+c_0 \bar{f}^{-c_1} e^{-c_2/\bar{f}}
  \label{eq:hcfit}
\end{equation}
where $a_i, b_i, c_i$ are constants to be determined by the fit and $\bar{f}=f/(10^{-8}{\rm Hz})$. We find that setting
\begin{align*}
a_0&= 7.27\times 10^{-14}\,\,\,\,\,\,\,\,\,\, & a_1&=0.254 & a_2&=0.807\\
b_0&= 1.853\times 10^{-12}\,\,\,\,\,\,\,\,\,\, & b_1&=1.77  & b_2&=3.7\\
c_0&= 1.12\times 10^{-13}\,\,\,\,\,\,\,\,\,\,   & c_1&=0.676 & c_2&=0.6
\end{align*}
reproduces the spectrum within a maximum error of 1.5\% in log-amplitude (i.e. 3.5\% in amplitude), as shown in figure \ref{fig:specfit}. It also shows the difference between the analytical fit presented in this paper versus \citep{HuertaEtAl:2015}. The lower frequency shape (left to the peak) is recovered more accurately by equation \eqref{eq:hcfit}.

With this fitting formula in hand, equation (\ref{eq:hshift}) readily enables the analytical evaluation of the spectrum for any desired pair of reference values $f_t$, $e_t=e(f_t)$ (note that those can be function of the MBHB parameters, e.g. its chirp mass, or of the environment in which the binary evolve, as we will see in Section \ref{sec:Coupling}). Moreover, equation (\ref{eq:hc0}) shows that the spectrum of a binary with different chirp mass and redshift can be simply obtained by multiplying $h_{c,{\rm fit}}(f)$ by $({\mathcal{M}}/{\mathcal{M}_0})^{5/3}$ and $(({1+z})/({1+z_0}))^{-1/3}$, respectively. Therefore, the overall spectrum of the MBHB population can be generated from $h_{c,{\rm fit}}(f)$ as
\begin{equation}
\begin{split}
  h_c^2(f) = & \int_{0}^{\infty} dz \int_{0}^{\infty} d{\cal M} \frac{d^2n}{dzd{\cal M}} h_{c,{\rm fit}}^2\Big(f\frac{f_{p,0}}{f_{p,t}}\Big) \\ & \Big(\frac{f_{p,t}}{f_{p,0}}\Big)^{-4/3} \Big(\frac{\mathcal{M}}{\mathcal{M}_0}\Big)^{5/3} \Big(\frac{1+z}{1+z_0}\Big)^{-1/3},
  \label{eq:hcanalytic}
\end{split}
\end{equation}
where the ratio $f_{p,0}/f_{p,t}$ is calculated by means of equation (\ref{eq:fpeak}).


\subsection{Range of applicability}
\label{sec:applicability}
The assumption behind the above derivation is that the dynamics of the MBHB is purely driven by GW emission, i.e., its evolution is defined by equations (\ref{eq:dfdt}) and (\ref{eq:dedt}) formally back to $f= -\infty$. This of course cannot be true in practice, the question is whether the derivation provides a good approximation in the frequency range relevant to PTA detection.

\begin{figure}
\includegraphics[width=0.45\textwidth]{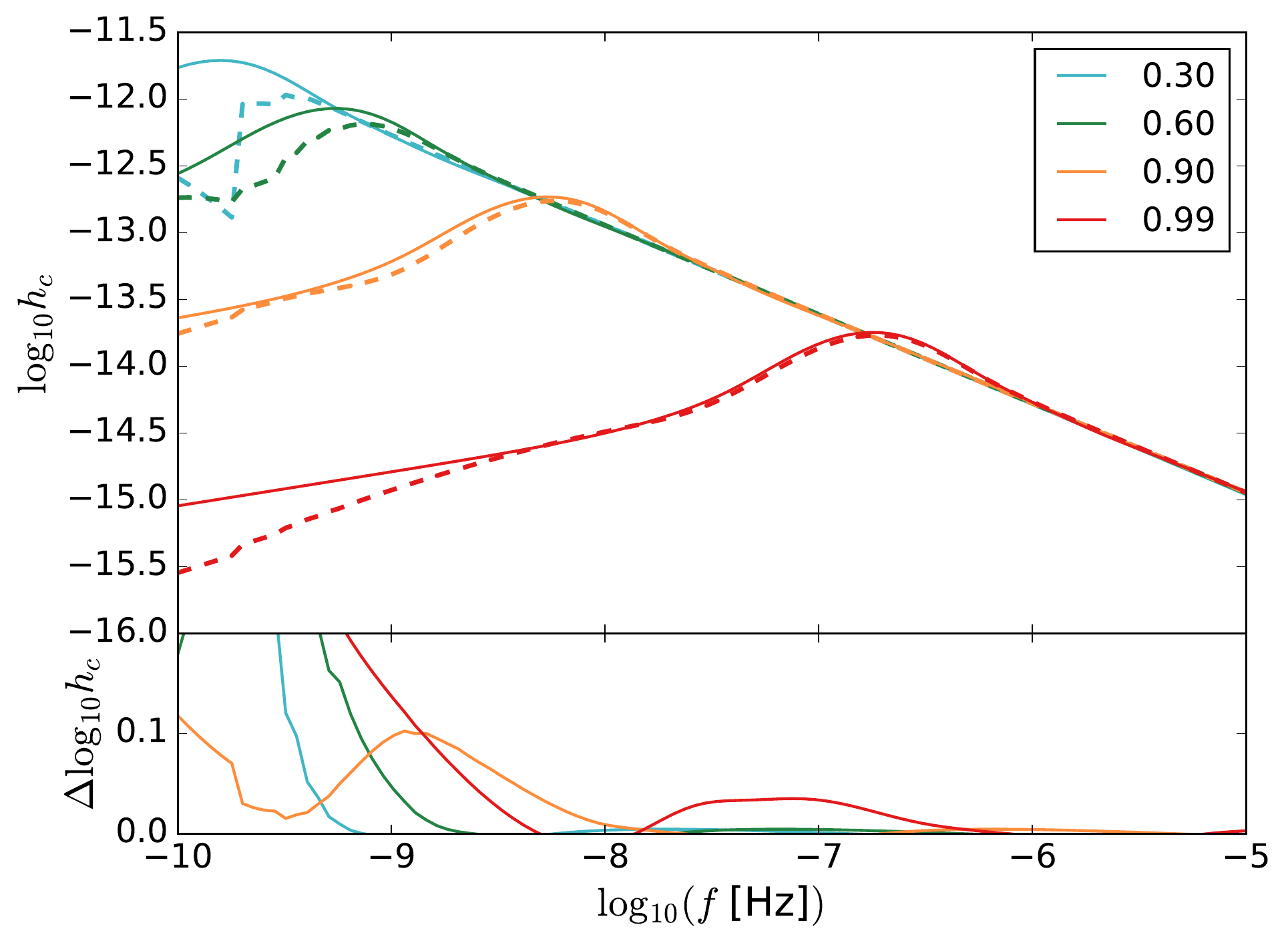}
\caption{characteristic amplitude spectrum for different eccentricities calculated with $n = 12500$ harmonics where only frequencies $f_n \geq 10^{-10}$ contribute (dashed lines) compared to the spectrum computed with no limitations on $f_n$ (solid lines). The lower panel shows the difference ${\rm log}_{10}h_{c,{\rm fit}}-{\rm log}_{10}h_{c}$ as a function of frequency for the different cases, and it is always $<0.1$ for $f>1\,$nHz.}
\label{fig:speccompare}
\end{figure}

MBHBs are driven by coupling with their environment up to a certain {\it transition orbital frequency}, $f_t$. At lower frequencies the evolution is faster than what is predicted by GW emission only and the eccentricity does not indefinitely grow to approach $e=1$. If the lowest frequency probed by PTA is $f_{\rm min}$ (which is $1/T$, where $T$ is the observation time, as defined in the introduction), then a necessary requirement for the applicability of equation (\ref{eq:hcanalytic}) is $f_t<f_{\rm min}$. This is, however, not a sufficient condition because for an eccentric MBHB population, the spectrum at $f_{\rm min}$ is defined by the contribution of binaries emitting at $f_n<f_{\rm min}$ satisfying the requirement $f_n=f_{\rm min}(1+z)/n$ for some $n$. If $f_t=f_{\rm min}$, and therefore the binary evolves faster and is less eccentric at $f_n<f_{\rm min}$, then the contribution of the $n$-th harmonics of systems emitting at $f_n$ is smaller, affecting the overall spectrum at $f_{\rm min}$ and above.

To investigate the impact of this fact on the spectrum we consider the same reference binaries with transition frequency $f_t=f_0=0.1$nHz and $e_t=e_0=0.3, 0.6, 0.9, 0.99$, but now assuming they {\it form} at $f_t$, i.e., discarding the contribution of lower frequencies to the computation of the spectrum. The result is compared to the full spectrum in figure \ref{fig:speccompare}. As expected, the absence of binaries at $f<f_t$ partially suppresses the signal observed at $f>f_t$. However three things should be noticed: i) the suppression is relevant only up to $f\sim 10f_t$, ii) the effect is small for highly eccentric binaries -- this is because for large $e$, the gravitational wave strain $h_c$ is dominated by the first, rather than the second harmonic, see figure 4 in \cite{2016ApJ...817...70T}--, and iii) this is the most pessimistic case, since for a realistic orbital evolution, binaries do emit also at $f<f_t$, but their contribution to the spectrum at $f>f_t$ is smaller due to the faster evolution and lower eccentricity. Therefore, our approximation should hold in the PTA band as long as the typical transition frequency $f_t$ is few time smaller than $f_{\rm min}$. In the next section we will show that for a typical MBHB population driven by scattering of stars this is indeed generally the case. 

\section{Binaries in stellar environments}
\label{sec:Coupling}

Following galaxy mergers, MBHBs sink to the centre because of dynamical friction \citep{1943ApJ....97..255C} eventually forming a bound pair when the mass in star and gas enclosed in their orbit is of the order of the binary mass. For MBHBs with $M=M_1+M_2>10^8\msun$ relevant to PTA, this occurs at an orbital separation of few parsecs, and the corresponding GW emission is well outside the PTA band. At this point, dynamical friction becomes highly inefficient, and further hardening of the binary proceeds via exchange of energy and angular momentum with the dense gaseous and stellar environment \citep[see][and references therein]{DottiSesanaDecarli:2012}. 

The bulk of the PTA GW signal is produced by MBHBs hosted in massive galaxy (generally spheroids) at redshift $<1$. \cite{Sesana:2013} and \cite{2015MNRAS.447.2772R} further showed that the vast majority of the signal comes from 'red' systems, featuring old stellar populations and only a modest amount of cold gas. This fact does not immediately imply that MBHBs cannot be driven by interaction with cold gas in a form of a massive circumbinary disk. After all, because of the observed MBH-host galaxy relations \citep[see, e.g.][]{KormendyHo:2013}, even a mere 1\% of the galaxy baryonic mass in cold gas is still much larger than the MBHB mass, and therefore sufficient to form a circumbinary disk with mass comparable to the binary, if concentrated in the very centre of the galaxy. On the other hand, the relative fraction of observed bright quasars declines dramatically at $z<1$ \citep[e.g.][]{2007ApJ...654..731H}, implying that accretion of large amounts of cold gas, and hence a scenario in which MBHBs evolve in massive circumbinary disks, is probably not the norm. We therefore concentrate here on MBHBs evolving via interaction with stars.

\cite{SesanaKhan:2015} have shown that, following the merger of two stellar bulges, the evolution of the bound MBHBs can be approximately described by the scattering experiment formalism developed by \cite{1996NewA....1...35Q}. In Quinlan's work, the binary semimajor axis evolution follows the simple equation
\begin{equation}
 \frac{da}{dt}=-\frac{HG\rho a^2}{\sigma},
  \label{eq:dadt}
\end{equation}
where $\rho$ is a fiducial stellar background density and $\sigma$ the characteristic value of the Maxwellian distribution describing the velocity of the stars. $H$ is a dimensionless constant (empirically determined by the scattering experiments) of order $15-20$, largely independent on the MBHB mass ratio $q=M_2/M_1$ and eccentricity $e$. \cite{SesanaKhan:2015} found that equation (\ref{eq:dadt}) is applicable to post merger stellar distributions providing that $\sigma$ is the typical velocity dispersion of the stellar bulge and $\rho$ is the average stellar density at the MBHB influence radius, $\rho_i=\rho(r_i)$, defined approximately as the radius enclosing a stellar mass twice the total MBHB mass $M=M_1+M_2$. In the stellar dynamic jargon, this corresponds to a situation where the MBHB 'loss-cone' is full at the binary influence radius. By using different methods, \cite{2015ApJ...810...49V} came to similar conclusions stressing, however, that in the long term the MBHB hardening rate tends to drop compared to equation (\ref{eq:dadt}), a hint that the loss-cone might not be kept full in the long term. The evolution of the Keplerian orbital frequency $f_K$ of the MBHB can therefore be written as:
\begin{equation}
  \frac{df_K}{dt}=\frac{df_K}{dt}\Big{|}_* + \frac{df_K}{dt}\Big{|}_{gw},
  \label{eq:fcombined}
 \end{equation}
where
\begin{align}
  \frac{df_K}{dt}\Big{|}_* & = \frac{3}{2 (2\pi)^{2/3}} \frac{H\rho_i}{\sigma} G^{4/3} M^{1/3} f_K^{1/3},
  \label{eq:fstar}
\\
\frac{df_K}{dt}\Big{|}_{gw} & = \frac{96}{5} (\pi)^{8/3} \frac{G^{5/3}}{c^5} \mathcal{M}^{5/3} f_K^{11/3} F(e).
  \label{eq:fgw}
\end{align}
Equation (\ref{eq:fstar}) is readily obtained from equation (\ref{eq:dadt}) by using Kepler's law, and equation (\ref{eq:fgw}) is the standard GW frequency evolution already seen in the previous section. It is easy to show that at low frequency stellar hardening dominates and GW takes over at a transition frequency that can be calculated by equating the two evolution terms to obtain:
\begin{equation}
\begin{split}
  f_{t}  & =
(2\pi)^{-1} \Big(\frac{5H\rho_i}{64\sigma F(e)}\Big)^{3/10} \frac{c^{3/2}}{G^{1/10}} \frac{(1+q)^{0.12}}{q^{0.06}} \mathcal{M}^{-2/5}\\
& \approx 0.56\pi^{-1} \Big(\frac{5H\rho_i}{64\sigma F(e)}\Big)^{3/10} \frac{c^{3/2}}{G^{1/10}} \mathcal{M}^{-2/5} \\
  & = 0.356\, {\rm nHz}\, \left(\frac{1}{F(e)}\frac{\rho_{i,100}}{\sigma_{200}}\right)^{3/10}\mathcal{M}_9^{-2/5}
\label{eq:ft}
\end{split}
\end{equation}
Where $\rho_{i,100}=\rho_i/(100\,\msun{\rm pc}^{-3})$, $\sigma_{200}=\sigma/(200\,{\rm km\,s}^{-1})$, $\mathcal{M}_9=\mathcal{M}/(10^9\,\msun)$ and we assume $H=16$ in the last line. We notice that in the mass ratio $0.1<q<1$, that by far dominates the PTA GW signal \citep[see, e.g. figure 1 in][]{2012MNRAS.420..860S}, the function $(1+q)^{0.12}/q^{0.06}$ falls in the range $[1.08,1.15]$. Therefore, in the last two lines of equation (\ref{eq:ft}) we neglected the mass ratio dependence by substituting $(1+q)^{0.12}/q^{0.06}=1.12$. 
A fair estimate of the MBHB coalescence timescale is provided by the evolution timescale at the transition frequency, $t_c=f_t(dt/df_t)$. By using equations (\ref{eq:ft}) and (\ref{eq:fstar}) one obtains

\begin{equation}
\begin{split}
  t_c & = \frac{5}{96} (2\pi)^{-8/3} \frac{c^5}{G^{5/3}} \mathcal{M}^{-5/3} f_t^{-8/3} F(e)^{-1}\\
  & = \frac{2}{3} \frac{c}{G^{7/5}} \Big(\frac{\sigma}{H\rho_i}\Big)^{4/5} \Big(\frac{5}{64F(e)}\Big)^{1/5} \frac{q^{0.16}}{(1+q)^{0.32}} \mathcal{M}^{-3/5}\\
  & \approx 0.5 \frac{c}{G^{7/5}} \Big(\frac{\sigma}{H\rho_i}\Big)^{4/5} \Big(\frac{5}{64F(e)}\Big)^{1/5} \mathcal{M}^{-3/5}\\
  & = 0.136 \ {\rm Gyr} \ F(e)^{-1/5}\left(\frac{\rho_{i,100}}{\sigma_{200}}\right)^{4/5}\mathcal{M}_9^{-3/5}
  \label{eq:tcoal}
\end{split}
\end{equation}
where, once again, we omitted mild $q$ dependences in the last approximation by substituting $q^{0.16}/(1+q)^{0.32}=0.75$ ($0.67 < q^{0.16}/(1+q)^{0.32} < 0.8$ for $0.1<q<1$).



For an operational definition of $f_t$ and $t_c$, we need to define $\rho_i$ and $\sigma$. The density profile of massive spheroidals is well captured by the Dehnen's density profile family \citep{Dehnen:1993} which takes the form
\begin{equation}
\rho(r) = \frac{(3-\gamma)M_* a}{4\pi} r^{-\gamma} (r+a)^{\gamma-4}
\end{equation}
where $0.5 < \gamma < 2$ determines the inner slope of the stellar density distribution, $M_*$ is the total mass of the bulge in star, $a$ is its characteristic radius. The influence radius $r_i$ of the MBHB is then set by the condition
\begin{equation}
  2M = \int_0^{r_i} 4\pi r^2 \rho(r) dr
  \label{eq:ricondition}
\end{equation}
which gives
\begin{equation}
  r_i = \frac{a}{(2M/M_*)^{1/(\gamma-3)}-1}.
\end{equation}
Inserting $r_i$ back into the Dehnen profile gives
\begin{equation}
\rho_i \approx \frac{(3-\gamma)M_*}{4\pi a^3} \Big(\frac{2M}{M_*}\Big)^{\gamma/(\gamma-3)}
\end{equation}
where we used the fact that $2M << M_*$. It is possible to reduce the number of effective parameters $M, M_*, a, \gamma, \sigma$ by employing empirical relations connecting pairs of them, valid for stellar spheroids. In particular we use the $a-M_*$ relation of \cite{DabringhausenHilkerKroupa:2008}, and the $M-\sigma$ and $M-M_*$ of \cite{KormendyHo:2013} relations:
\begin{align}
  a & = 239 \,\text{pc}\, (2^{1/(3-\gamma)}-1) \Big(\frac{M_*}{10^9\msun}\Big)^{0.596}
  \label{eq:ascale}
  \\
  \sigma & = 261\, \text{km s}^{-1}\, \Big(\frac{M}{10^9\msun}\Big)^{0.228}
  \label{eq:msigma}
  \\
  M_* & = 1.84\times 10^{11}\,\msun \, \Big(\frac{M}{10^9\msun}\Big)^{0.862}.
  \label{eq:mbulge}
\end{align}
This allows to express $\rho_i$ as a function of $M$ and $\gamma$ only in the form

\begin{equation}
  \rho_i = 0.092 \,\msun\text{pc}^{-3} \,{\cal F}(\gamma) \Big(\frac{M}{10^9 M_\odot}\Big)^{{\cal G}(\gamma)},
  \label{eq:rhoi}
\end{equation}
where
\begin{align}
  {\cal F}(\gamma) & = \frac{(3-\gamma) 92^{\gamma/(3-\gamma)}}{(2^{1/(3-\gamma)}-1)^3}\nonumber\\
  {\cal G}(\gamma) & = -0.68-0.138\frac{\gamma}{3-\gamma}\nonumber
\end{align}
Equations (\ref{eq:msigma}) and (\ref{eq:rhoi}) are expressed as a function of $M$. However, we notice from equation (\ref{eq:ft}) that $f_t\propto M^{-0.3-0.041\gamma/(3-\gamma)}$. Since ${\cal M}=Mq^{3/5}/(1+q)^{6/5}$, if $0.1<q<1$, then $2.32 {\cal M}<M< 3.57{\cal M}$. It is easy to show that for $0.5<\gamma<2$, by substituting $M=2.9{\cal M}$, equation (\ref{eq:ft}) returns $f_t$ within 10\% of the correct value when $0.1<q<1$.

Finally, equation (\ref{eq:fcombined}) defines only the frequency evolution of the MBHB. For a complete description of the system, tracking of the eccentricity evolution is also required. Both scattering experiments and N-body simulations have shown that MBHB-star interactions tend to increase $e$. The increase is generally mild for equal mass binaries and the eccentricity at the transition frequency largely depends on the initial eccentricity at the moment of binary pairing. Because of this mild evolution at $f_K<f_t$ and in order to keep the problem simple, we approximate the eccentricity evolution of the MBHB as:
\begin{equation}
\frac{de}{dt}=
\begin{cases}
0\,\,\,\,\,\,\,\, {\rm if}\,\,\,\,\,\,\,\, f_K<f_t\\
-\frac{1}{15} (2\pi)^{8/3} \frac{G^{5/3}}{c^5} \mathcal{M}^{5/3} f_K^{8/3} G(e)\,\,\,\,\,\,\,\, {\rm if}\,\,\,\,\,\,\,\, f_K>f_t
  \label{eq:ecombined}
\end{cases}
\end{equation}

\section{Results: Gravitational wave spectra calculation}
\label{sec:Population}
\subsection{Dynamics of MBHBs: transition frequency and coalescence time}
\label{sec:fttcsingle}

\begin{figure*}
\centering
  \includegraphics[width=0.85\columnwidth,clip=true,angle=0]{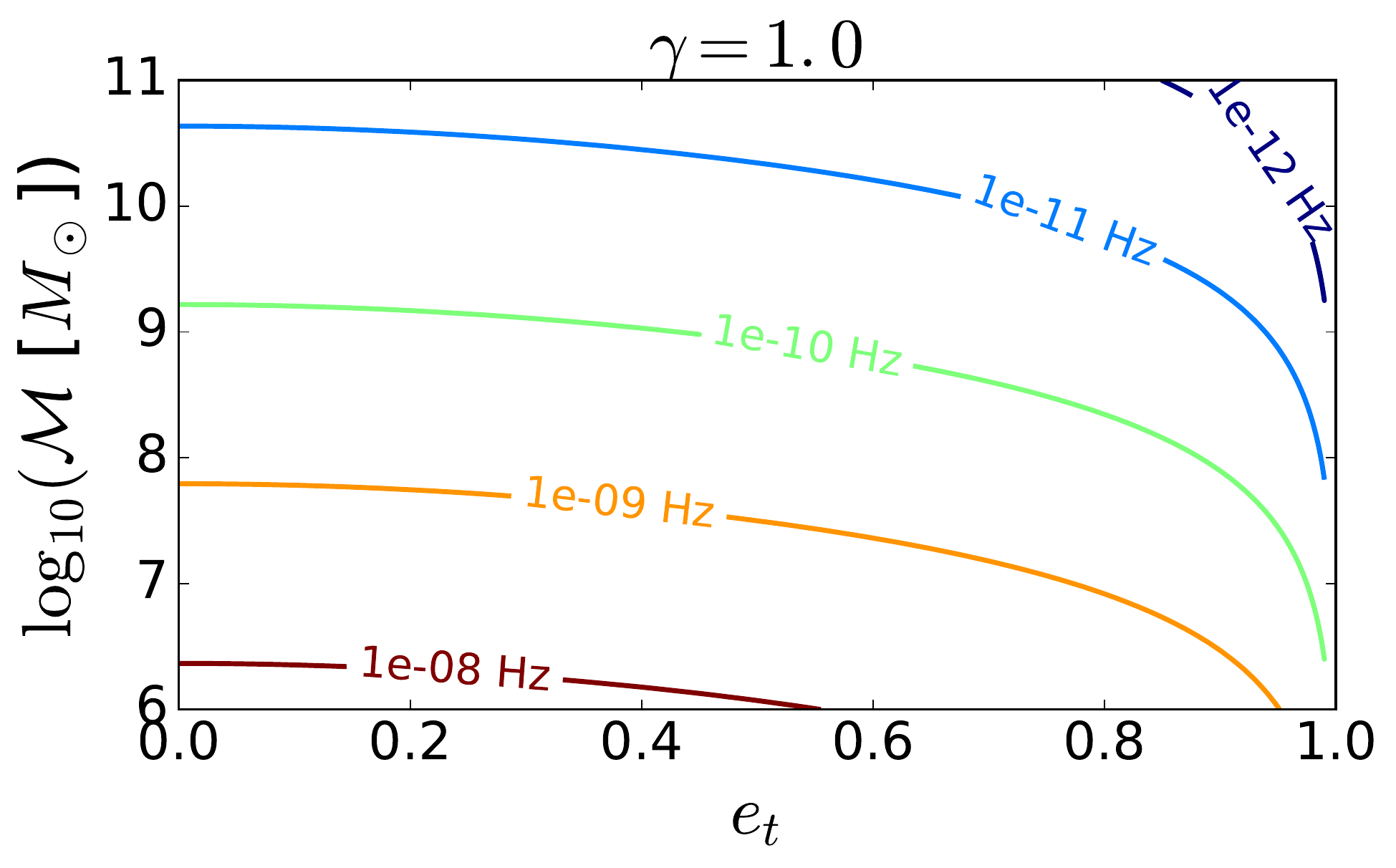}\hspace{1.5cm}
  \includegraphics[width=0.85\columnwidth,clip=true,angle=0]{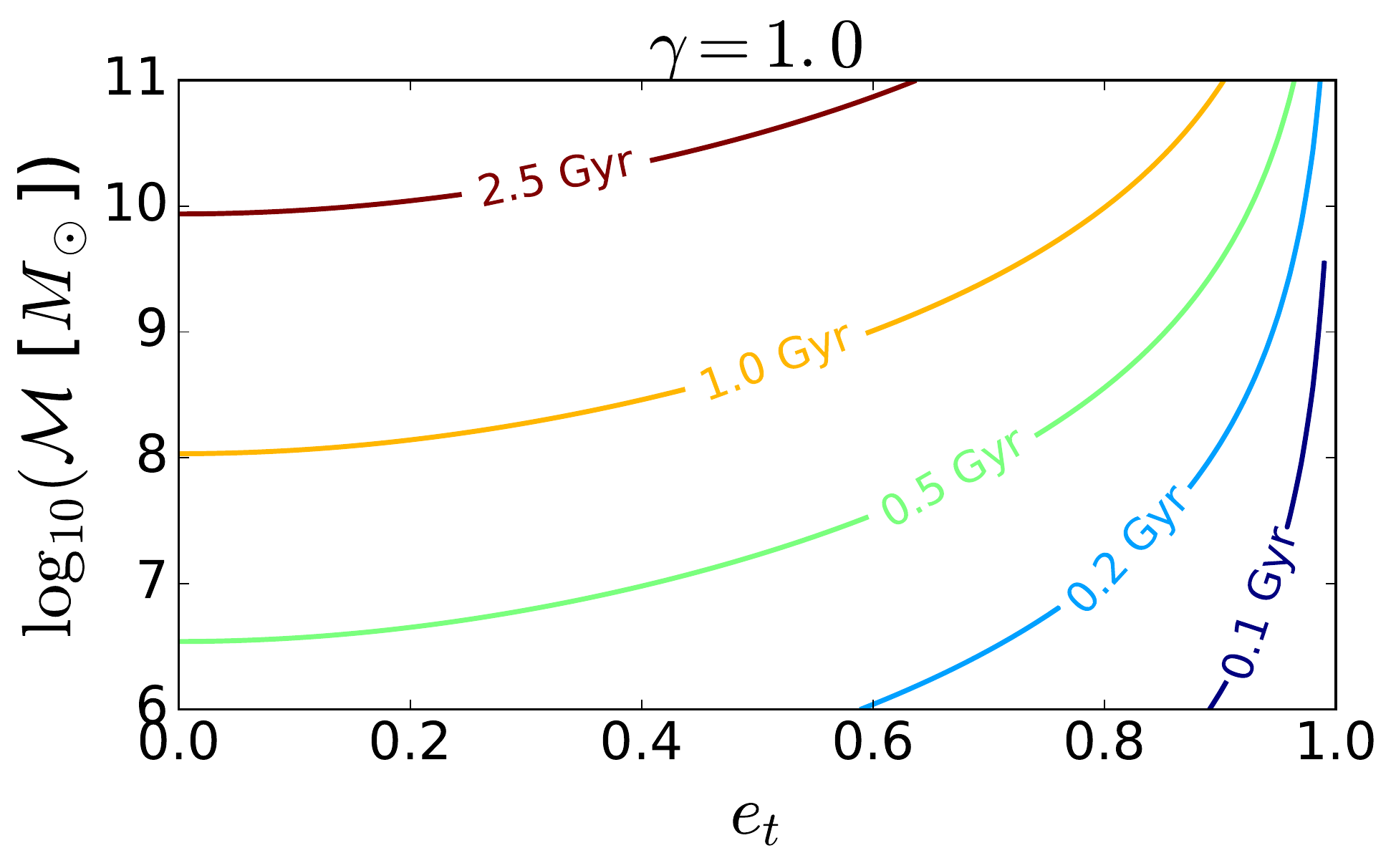}\\
  \includegraphics[width=0.85\columnwidth,clip=true,angle=0]{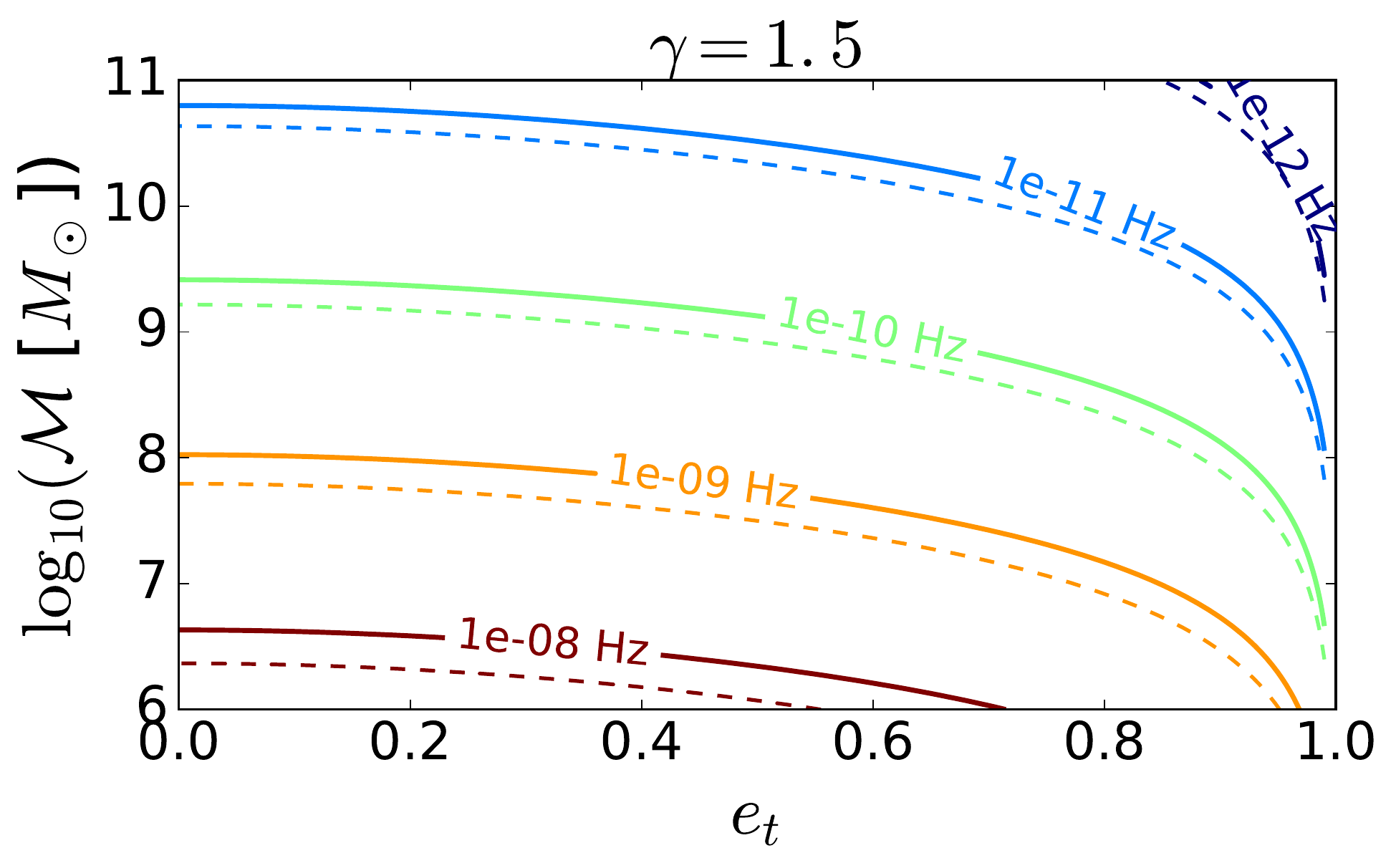}\hspace{1.5cm}
  \includegraphics[width=0.85\columnwidth,clip=true,angle=0]{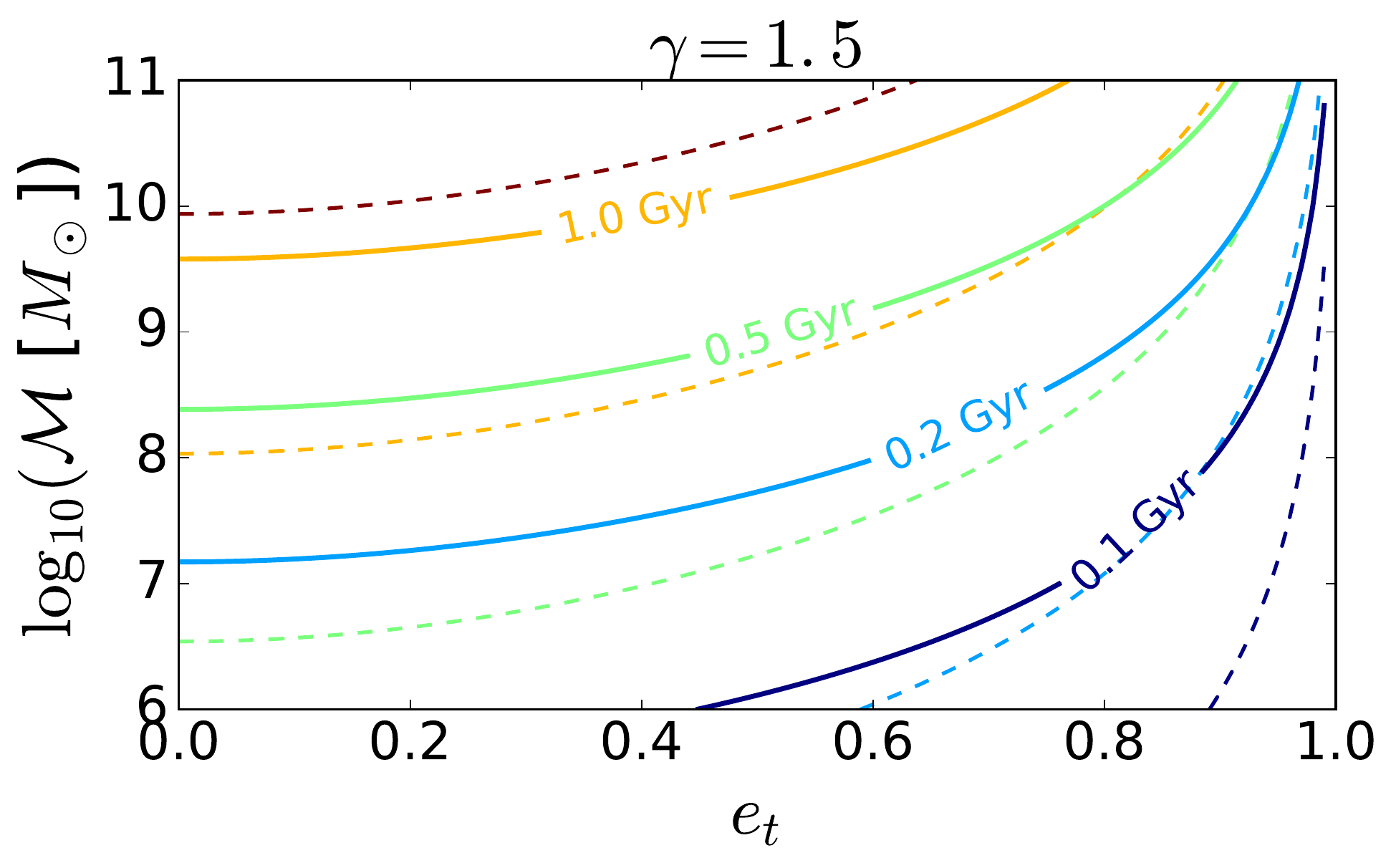}\\
  \includegraphics[width=0.85\columnwidth,clip=true,angle=0]{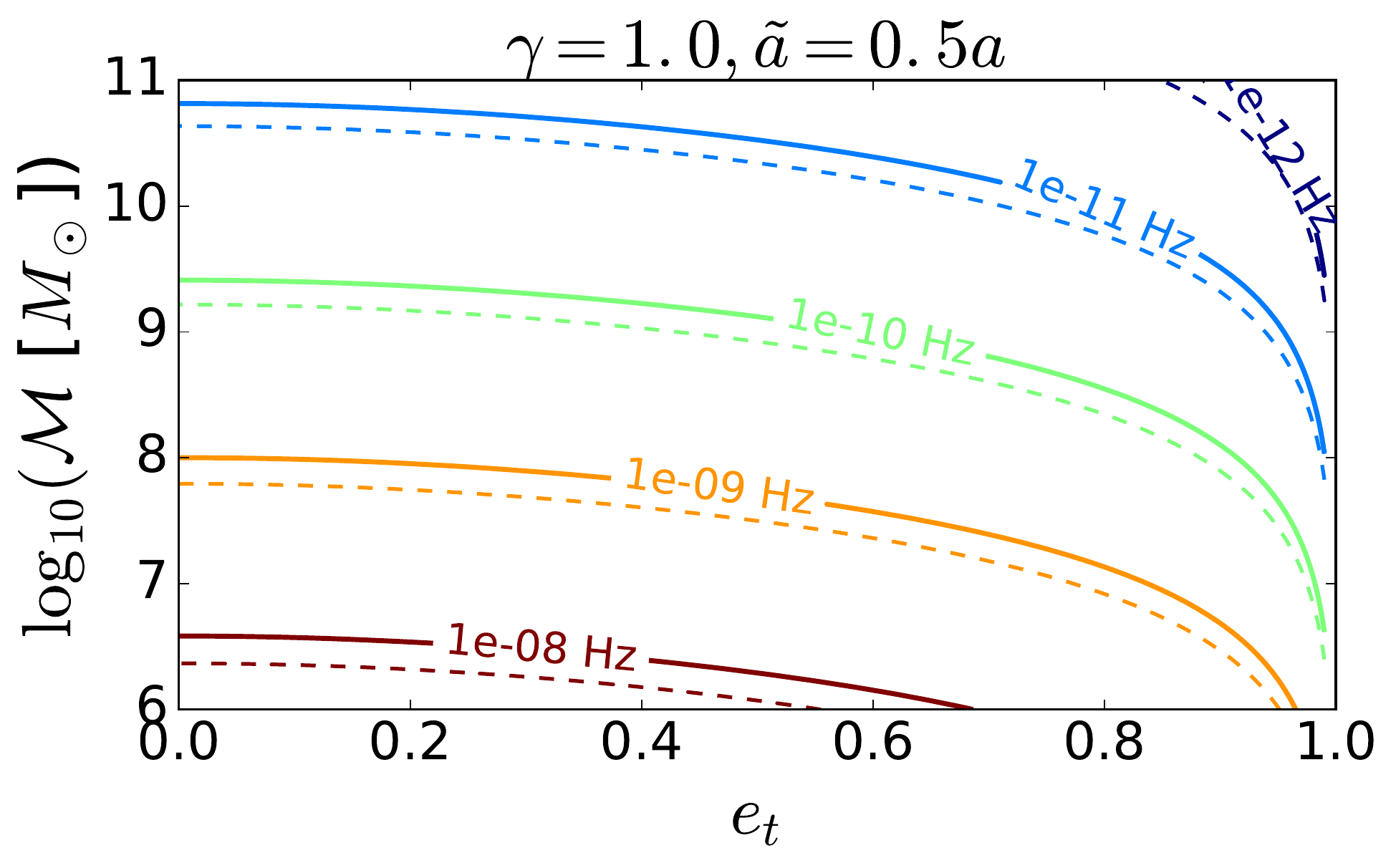}\hspace{1.5cm}
  \includegraphics[width=0.85\columnwidth,clip=true,angle=0]{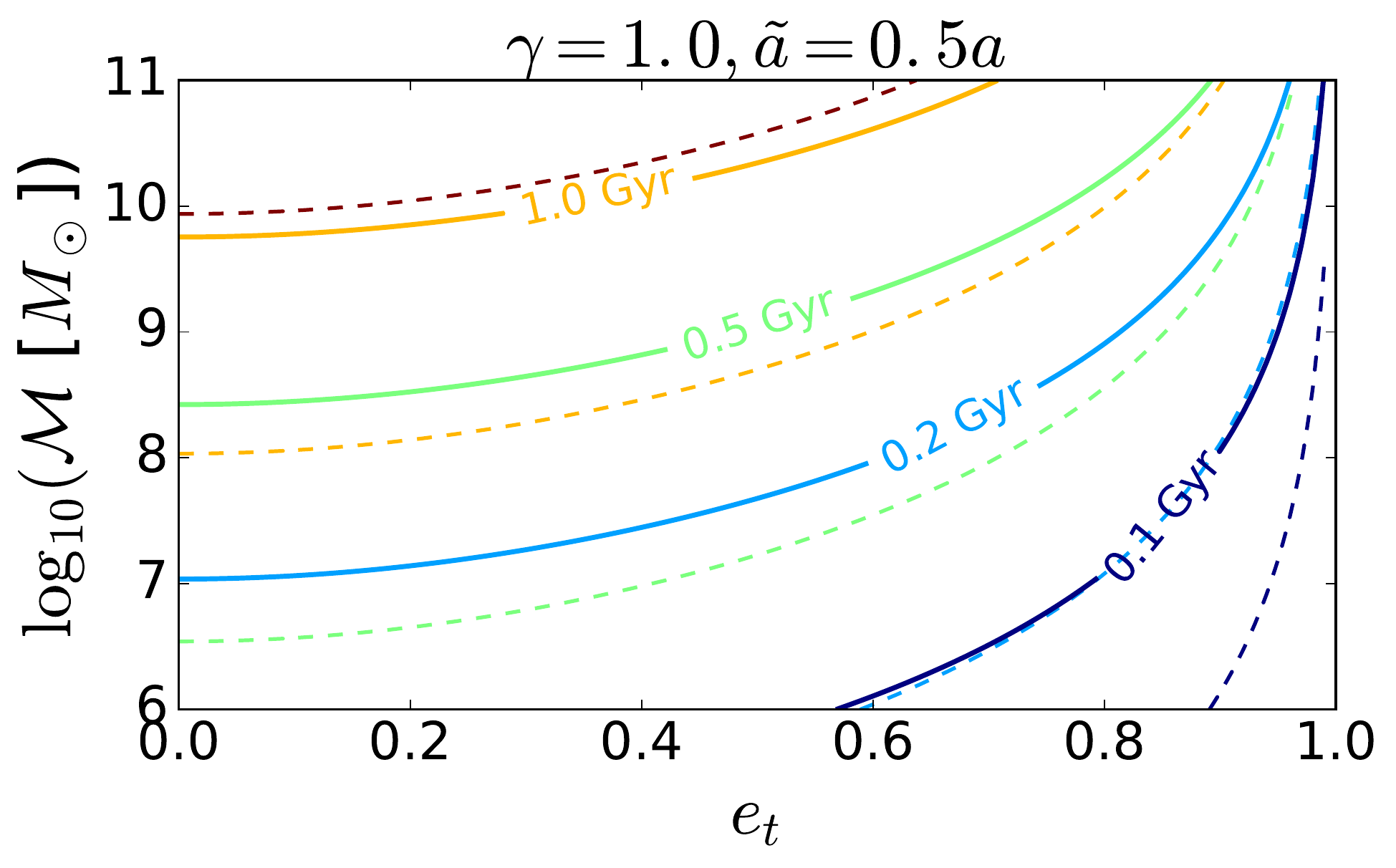}\\
  \includegraphics[width=0.85\columnwidth,clip=true,angle=0]{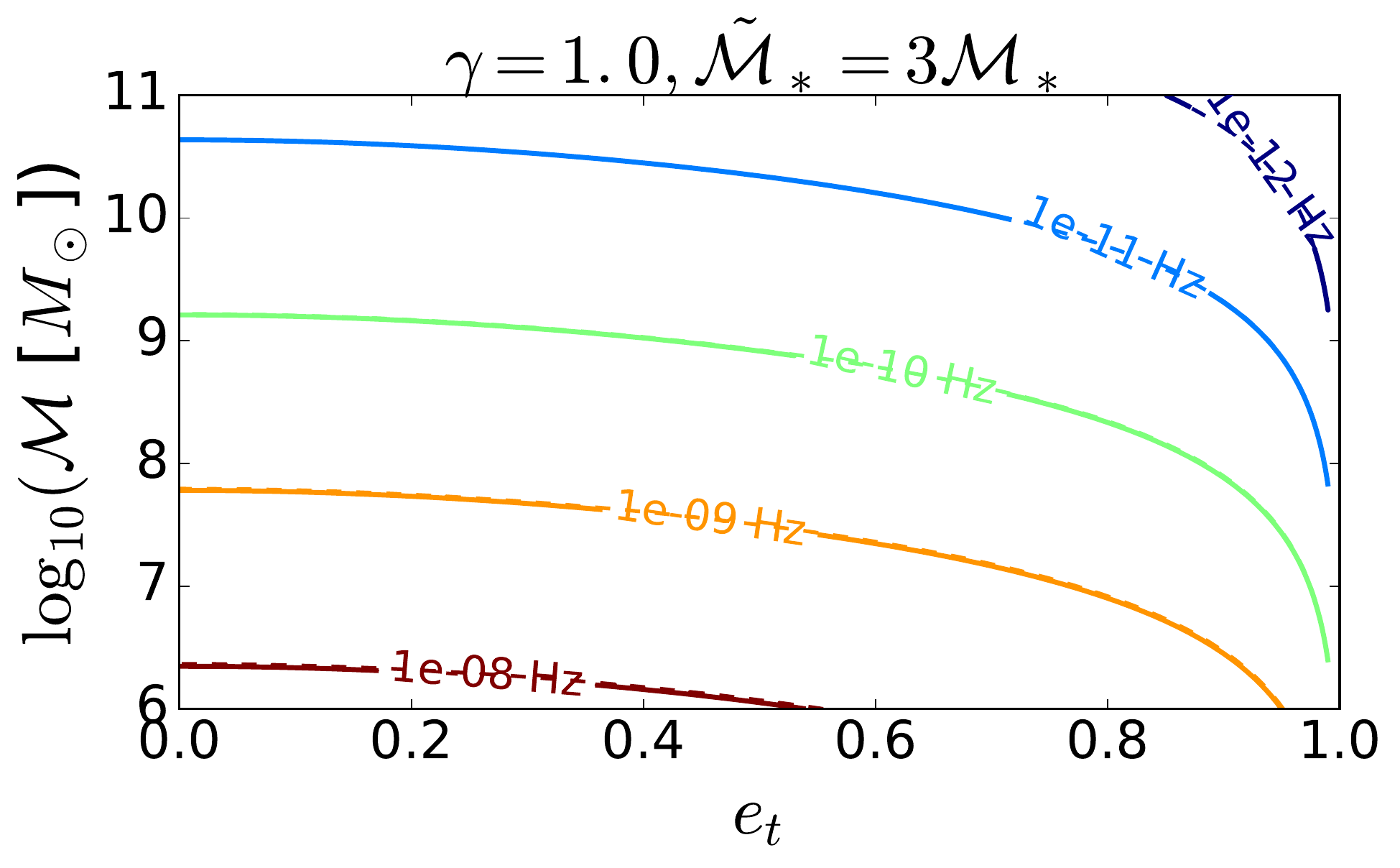}\hspace{1.5cm}
  \includegraphics[width=0.85\columnwidth,clip=true,angle=0]{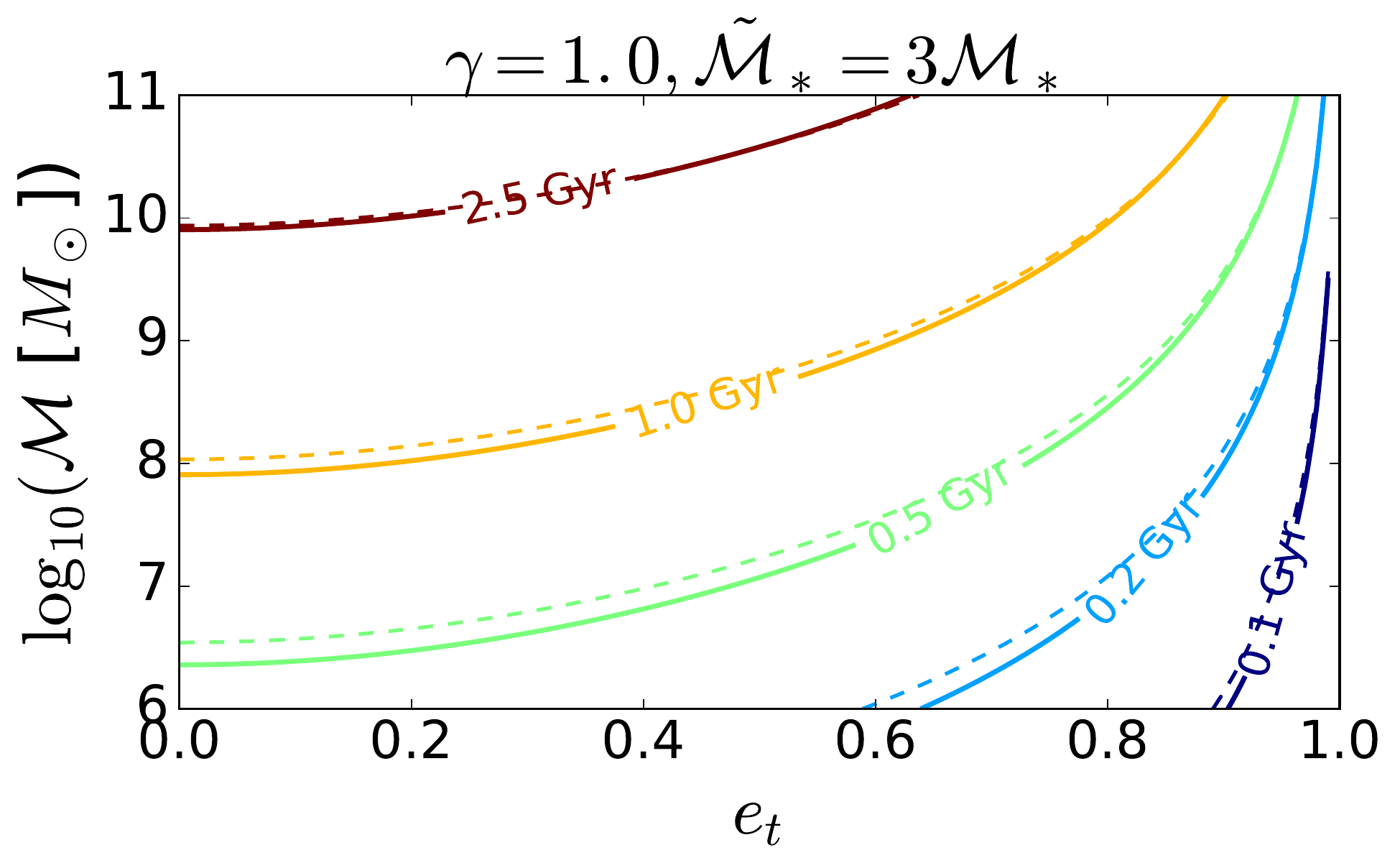}\\
  \caption{Contour plots of transition frequency $f_t$ (left panels) and coalescence timescales $t_c$ (right panels) in the transition eccentricity $e_t$ vs chirp mass $\mathcal{M}$ plane. $f_t$ is computed according to equation (\ref{eq:ft}) with $\rho_i$ provided by equation (\ref{eq:rhoi}) calculated assuming the fiducial scaling relations given by equations (\ref{eq:ascale},\ref{eq:msigma},\ref{eq:mbulge}). Shown are models with $\gamma = 1$ (which is our fiducial model, top row), $\gamma = 1.5$ (second row), $\gamma=1$ but $a$ decreased by a factor of two in equation (\ref{eq:ascale}) (third row) $\gamma=1$ and $M_*$ increased by a factor of three in equation (\ref{eq:mbulge}) (bottom row). The fiducial model is plotted in each of the other panels with dashed lines to highlight changes in $f_t$ and $t_c$ against the other models (plotted as solid lines).}
  \label{fig:ftandtc}
\end{figure*}

Before going into the computation of the GW spectrum, we can have a look at how transition frequency $f_t$ and coalescence timescale $t_c$ change as a function of ${\cal M}$ and $e_t$. In the following we consider four selected models representative of a range of physical possibilities having a major impact on the MBHB dynamics. Results are shown in figure \ref{fig:ftandtc}. The top panel shows a model with $\gamma=1$ and $\rho_i$ given by equation (\ref{eq:rhoi}). We consider this as our default model, because most of the PTA signal is expected to come from MBHBs hosted in massive elliptical galaxies with relatively shallow density profiles.
The GW signal is generally dominated by MBHBs with ${\cal M}>3\times 10^8\msun$, which are therefore our main focus. At low $e_t$ those systems have $f_t<0.3$nHz and coalescence timescales in the range $1.5-4$Gyr. For $e_t=0.9$, $f_t$ is ten times lower, nonetheless $t_c$ is roughly an order of magnitude shorter, in virtue of the $F(e)$ factor appearing in equation (\ref{eq:tcoal}). The effect of a steeper density profile is shown in the second row of plots in figure \ref{fig:ftandtc}, where we now assume $\gamma=1.5$. The effect of a steeper inner power law, is to make the stellar distribution more centrally concentrated, thus enhancing $\rho_i$. This makes stellar hardening more efficient and shifts $f_t$ by a factor $\approx 1.3$ upwards making $t_c$ a factor of $\approx 2$ shorter (using a shallower profile $\gamma=0.5$ would have an opposite effect of the same magnitude). We recognize that $\rho_i$ given by equation (\ref{eq:rhoi}) relies on a number of scaling relations that are constructed on a limited sample of local, non-merging, galaxies. We therefore also explore the effect of a bias in some of those relations. For example, merging galaxies might be more centrally concentrated and we explore this possibility by arbitrarily reducing the typical scale radius $a$ by a factor of two compared to equation  (\ref{eq:ascale}). The effect is shown in the third row of panels of figure \ref{fig:ftandtc} assuming $\gamma=1$, and it is very similar (slightly larger) to the effect of the steeper ($\gamma=1.5$) density profile shown in the second row. Finally, it has been proposed that the MBH-galaxy relations might be biased high because of selection effects in the targeted galaxy samples. \cite{2016MNRAS.460.3119S} propose that the typical MBH mass might be in fact a factor $\approx 3$ lower than what is implied by equations (\ref{eq:msigma}) and (\ref{eq:mbulge}). We therefore explore a model featuring $\gamma=1$ but with MBH mass decreased by a factor of three for given galaxy properties. Results are shown in the bottom panels of figure \ref{fig:ftandtc}. For a given MBHB mass, this model implies just a minor change in $\rho_i$ and $\sigma$, with negligible effects of $f_t$ and $t_c$, compared to the fiducial model.


\subsection{GW spectra of fiducial MBHBs}

\begin{figure*}
\centering
\begin{tabular}{ccc}
  \includegraphics[width=0.95\columnwidth,clip=true,angle=0]{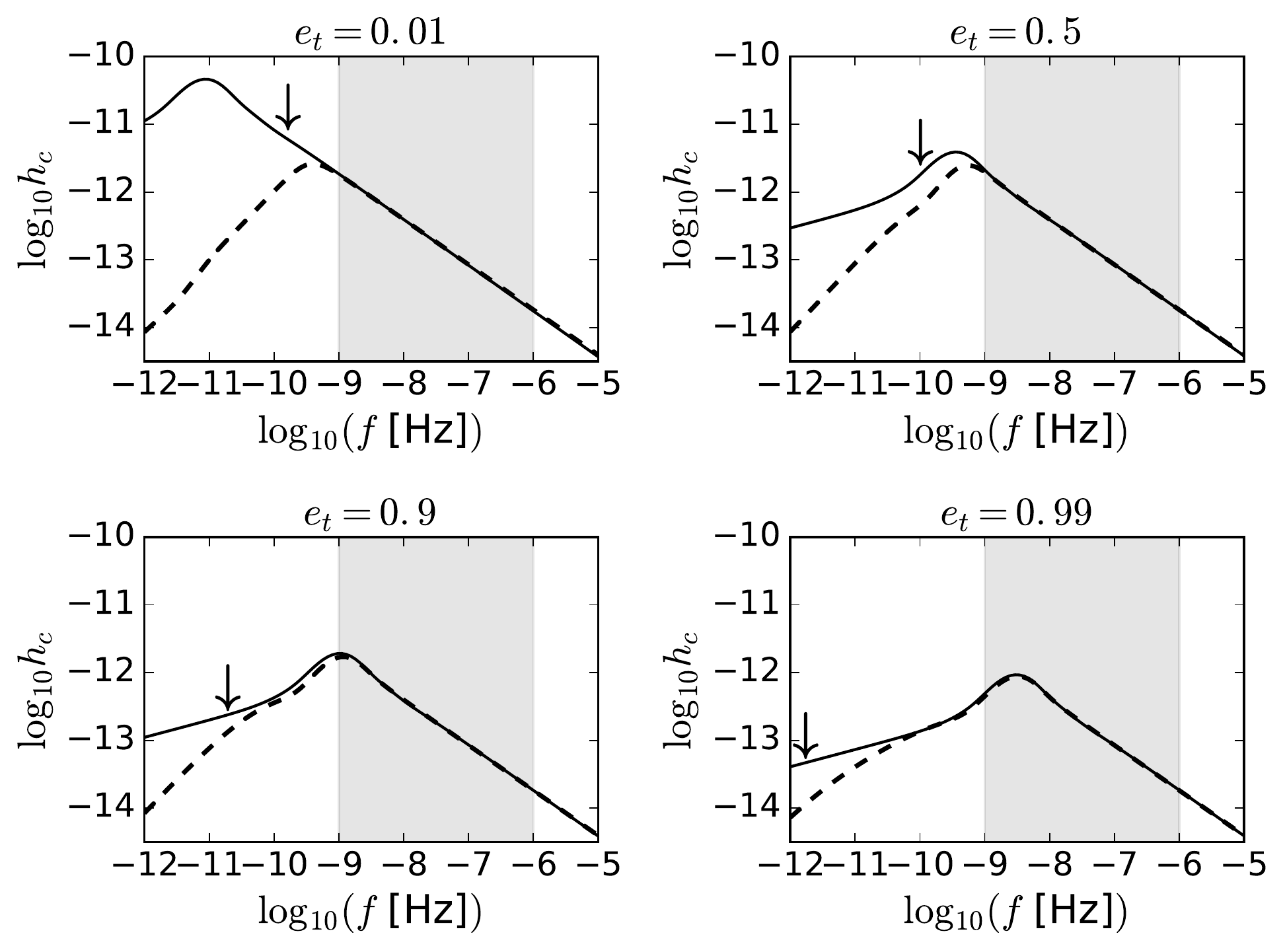}&\hspace{1.5cm}&
  \includegraphics[width=0.95\columnwidth,clip=true,angle=0]{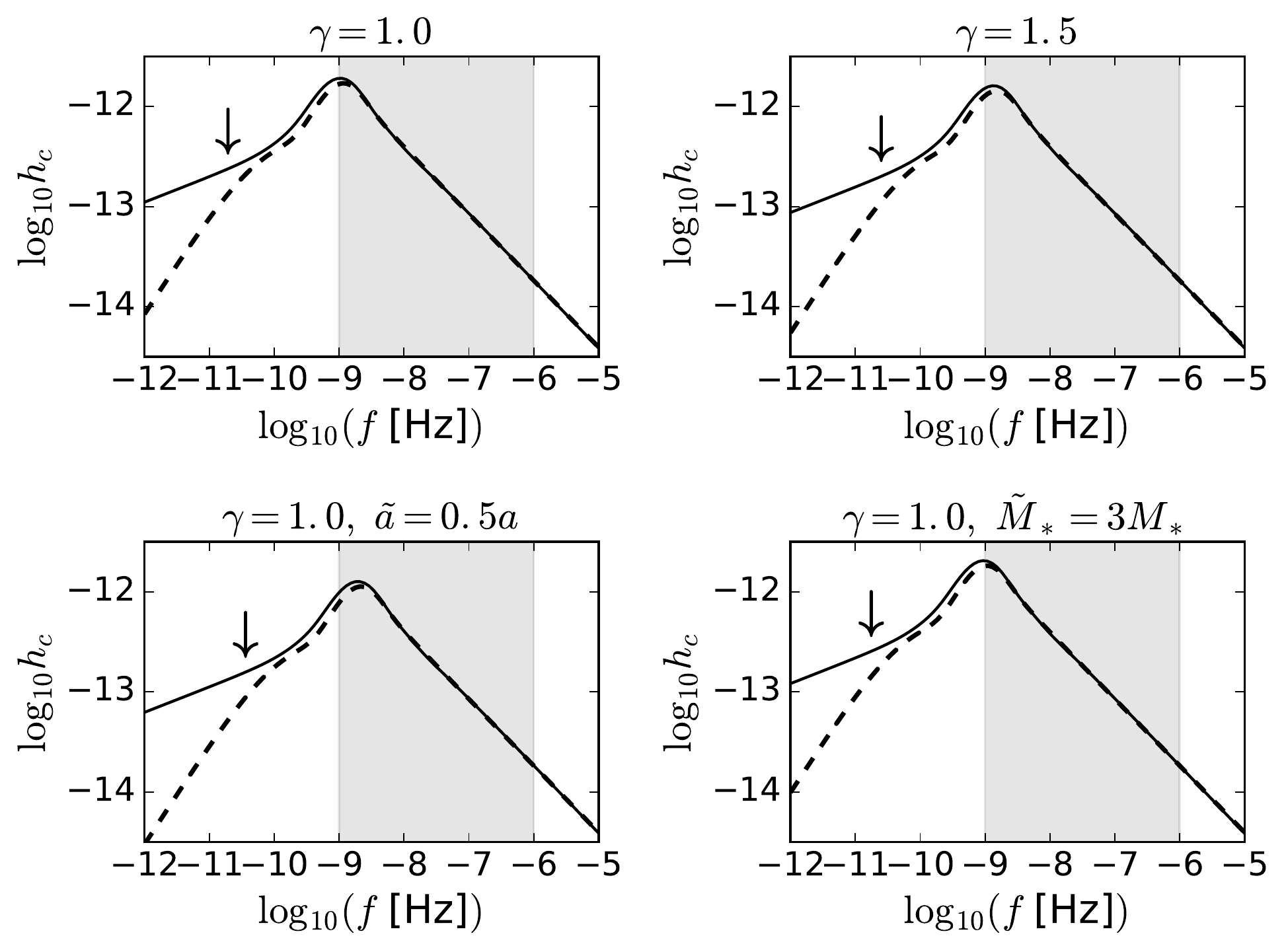}
\end{tabular}
  \caption{Spectra of a single MBHB with $\mathcal{M} = 10^9 M_\odot$, $z=0.02$. In each panel the solid line represents the spectrum fitted with equation (\ref{eq:hcfit}), whereas the dashed line considers the full evolution due to interaction with the stellar environment and eccentricity; the arrow indicates the turnover frequency. The left four panels are for our fiducial galaxy model ($\gamma = 1$) and different MBHB eccentricities as indicated in each panel; the right four panels are for $e_t = 0.9$, but assume different prescriptions of the astrophysical relations used to calculate the turnover frequency (i.e. assume different host galaxy models), as specified in each panel. The shaded stripe at $10^{-9}$Hz$<f<10^{-6}$Hz indicates the range of frequency relevant to PTA observations.}
  \label{fig:specsingle}
\end{figure*}

The GW spectrum generated by a MBHB evolving in a fiducial stellar background can now be computed by evaluating $dE/df_r$ in equation (\ref{eq:dEdf}), where the frequency and eccentricity evolution of the pair are now given by equations (\ref{eq:fcombined}) and (\ref{eq:ecombined}), instead of equations (\ref{eq:dfdt}) and (\ref{eq:dedt}), and the system is defined by the transition frequency $f_t$ as given in equation (\ref{eq:ft}) at which $e_t$ must be specified. We consider a fiducial MBHB with ${\cal{M}}=10^9\msun$  at $z=0.02$ ,and compare the real spectrum including stellar scattering to our approximated formula given by equation (\ref{eq:hcfit}) and appropriately re-scaled as described in section \ref{sec:fit}.

Results are shown in figure \ref{fig:specsingle} for all the environment models of figure \ref{fig:ftandtc}. In this and the following plots, solid lines are spectra computed via equation \eqref{eq:hcfit}, whereas dashed lines are spectra that includes stellar scattering driving the binary evolution at low frequency. We start by discussing the outcome of our fiducial model with $\gamma=1$, as a function of $e_t$, which is shown in the left plot. For circular binaries $f_t\approx0.2$nHz, well below the minimum PTA frequency $f_{\rm min}=1$nHz, appropriate for an PTA baseline of 30yrs, achievable within 2030. By increasing $e_t$, $f_t$ is pushed at lower values, eventually becoming irrelevant. Obviously, the real spectrum diverges from our analytic fit at $f<f_t$. Moreover, for moderately eccentric binaries ($e_t=0.5$ panel) the two spectra differ significantly up to almost $f=1$ nHz. This is mostly because the presence of the stellar environment 'freezes' the eccentricity to 0.5 at $f<f_t$; the real spectrum at $f\gtrsim f_t$ is missing the contribution from the very eccentric phase at $f<f_t$ that occurs when the environment is not taken into account and the binary is evolved back in time assuming GW emission only. The problem becomes less severe for larger values of $e_t$. Even though the presence of the environment freezes the binary eccentricity, $e_t$ is large enough that most of the relevant contribution from the higher harmonics emitted at low frequencies is kept. Most importantly, in all cases, at all $f>f_{\rm min}=1$nHz, our analytical fit perfectly describes the emitted spectrum. The right plot in figure \ref{fig:specsingle} shows the spectrum assuming $e_t=0.9$ for the four different environment models outlined in the previous subsection. Again, we notice that in all cases the GW spectrum is well described by our fitting formula in the relevant PTA frequency range, and the peak of the spectrum is only mildly affected (within a factor of two) by the different host models.



\subsection{Stochastic background from a cosmic MBHB population}


Having studied the signal generated by a fiducial system, we turn now to the computation of the overall GW spectrum expected from a cosmological population of MBHBs. To do this, we simply need to specify the distribution $d^2n/dzd{\cal M}$. We consider two population models:
\begin{itemize}
\item {\it model-NUM}: the $d^2n/dzd{\cal M}$ population is numerically constructed on the basis of a semi-analytic galaxy formation model implemented on the Millennium simulation \citep{SpringelEtAl_MilleniumSim:2005}, as described in \cite{SesanaVecchioVolonteri:2009}. In particular, we use a model implementing the $M_{\rm BH}-M_{\rm bulge}$ relation of \cite{2003ApJ...589L..21M}, with accretion occurring {\it before} the final MBHB coalescence on both MBHs in the pair.
\item {\it model-AN}: employs a parametrised population function of the form \cite{MiddletonEtAl:2016}
\begin{equation}
\frac{d^2 n}{dz d \log_{10} \mathcal{M}}  = \dot{n_0} \Big(\frac{\mathcal{M}}{10^7 M_\odot}\Big)^{-\alpha} e^{-\mathcal{M}/\mathcal{M}_*} (1+z)^\beta e^{-z/z_*} \frac{dt_r}{dz} \label{eq:pop},
\end{equation}
where $t_r$ is time measured in the source reference frame and 
\begin{equation}
  \frac{dt_r}{dz}  = \frac{1}{H_0 (1+z) (\Omega_M(1+z)^3 + \Omega_k(1+z)^2 + \Omega_\Lambda)^{1/2}}
  \label{eq:dtrdz}
\end{equation}
Based on loose cosmological constraints \citep[see][for details]{MiddletonEtAl:2016}, parameters lie in the range $\dot{n}_0 \in [10^{-20},10^3] \ \text{Mpc}^{-3} \text{Gyr}^{-1}, \ \alpha \in [-3,3], \ \mathcal{M}_* \in [10^6,10^{11}] \ M_\odot, \ \beta \in [-2,7], \ z_* \in [0.2,5]$. $H_0 = 70 {\,\text{km}}\,{\text{Mpc}^{-1}\text{s}^{-1}}$ is the Hubble constant and $\Omega_M = 0.3, \ \Omega_k = 0, \ \Omega_\Lambda = 0.7$ are the cosmological energy density ratios. We specialize our calculation to a fiducial mass function with $\log_{10}  \dot{n}_0 = -4, \ \alpha = 0, \ \mathcal{M}_* = 10^8 M_\odot, \ \beta = 2, \ z_* = 2$.
\end{itemize}


\begin{figure}
  \centering
  \includegraphics[width=0.95\columnwidth,clip=true,angle=0]{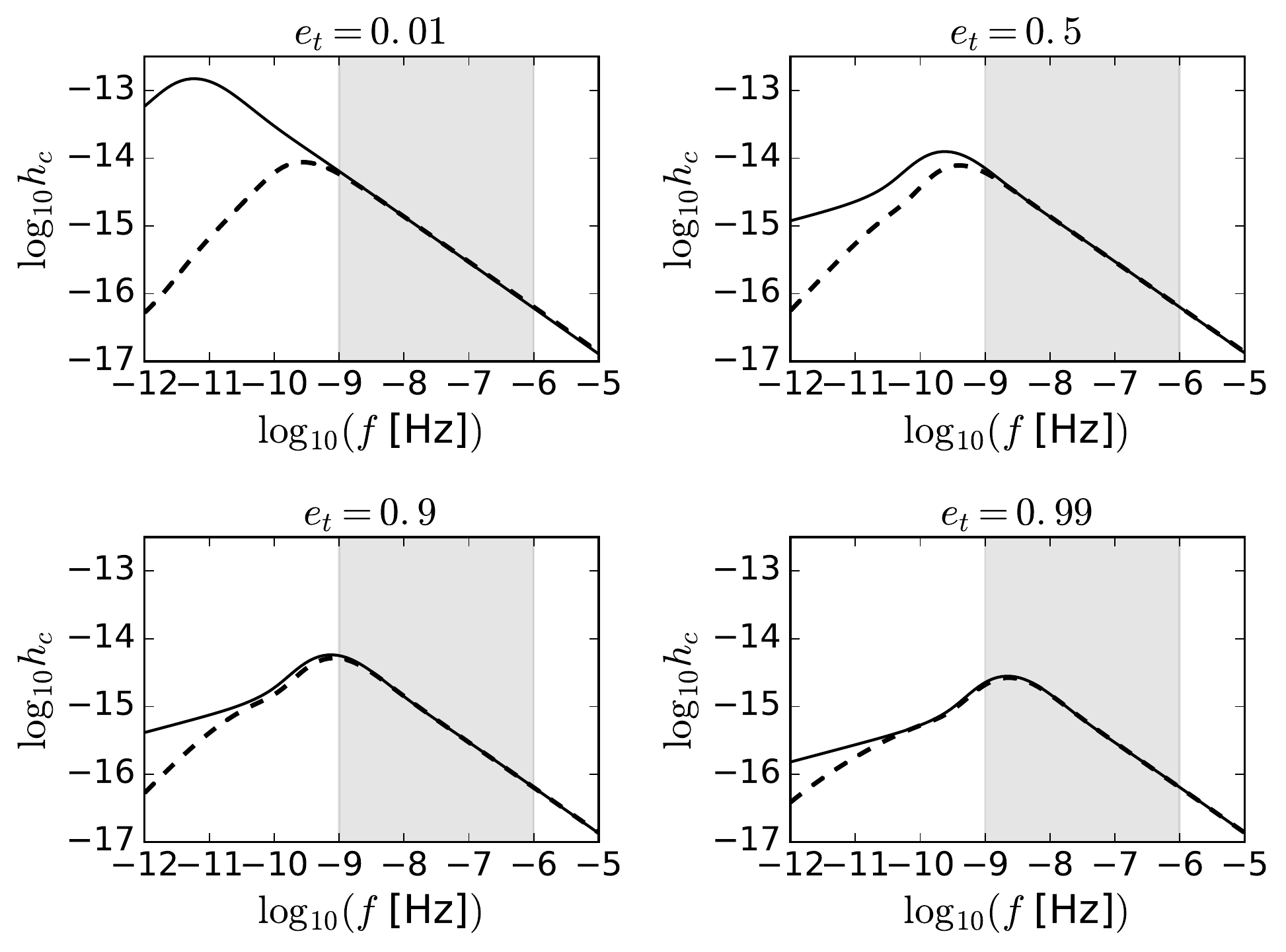}
  \caption{Same as the left half of figure \ref{fig:specsingle}, but the signal has now been integrated over the {\it model-NUM} MBHB population described in the text.}
  \label{fig:specpopnum}
\end{figure}

\begin{figure*}
\begin{subfigure}{0.45\textwidth}
\centering
$model-NUM$
\includegraphics[width=0.95\columnwidth]{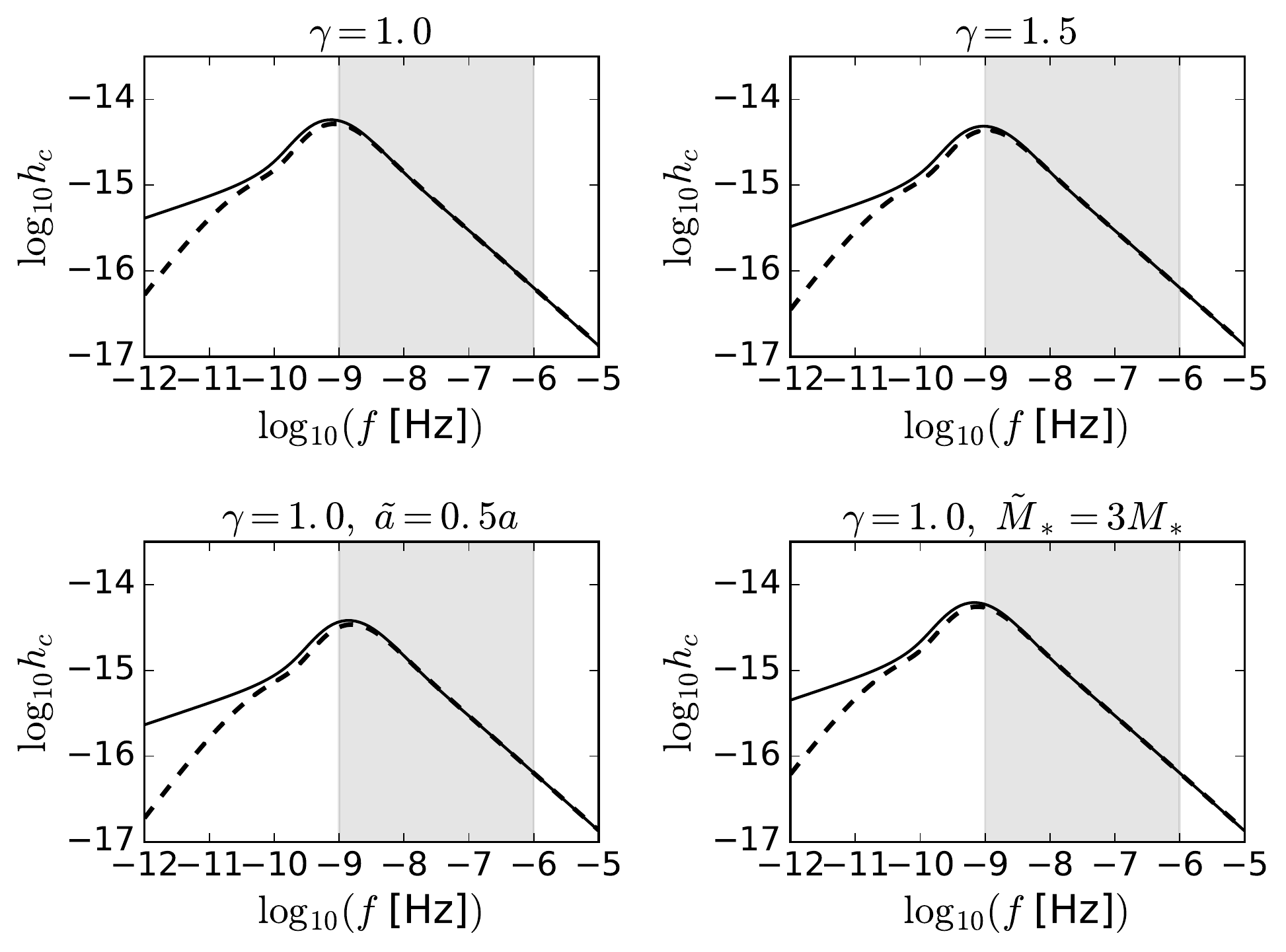}
\end{subfigure} \hspace{0.5cm}
\begin{subfigure}{0.45\textwidth}
\centering
$model-AN$
\includegraphics[width=0.95\columnwidth]{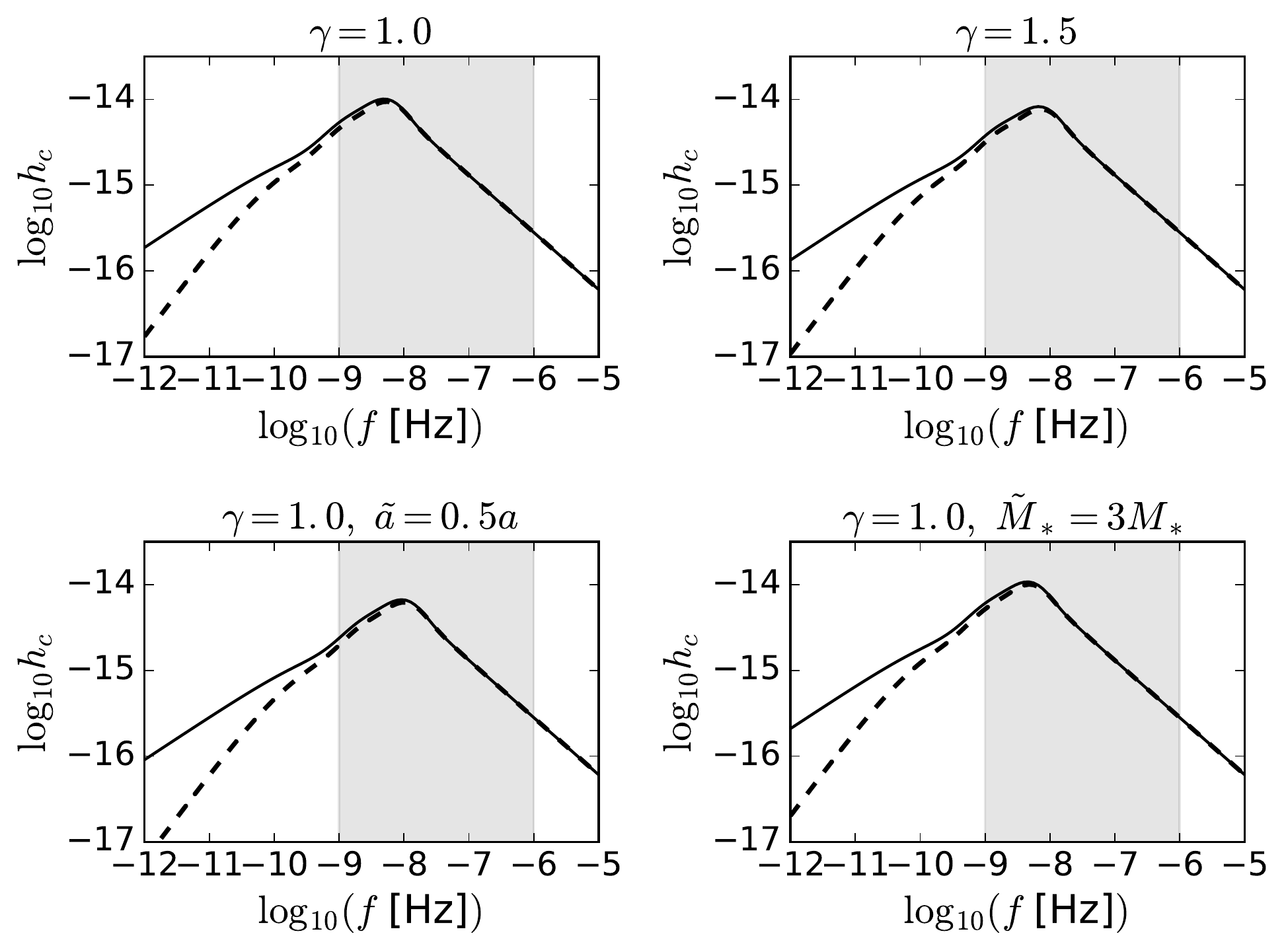}
\end{subfigure} \\
\caption{Same as the right half of figure \ref{fig:specsingle}, but the signal has now been integrated over the {\it model-NUM} (four panels on the left) and  {\it model-AN} (four panels on the right).}
\label{fig:specpopcomp}
\end{figure*}


To construct the spectrum we still need to specify a reference eccentricity $e_t$ at a reference binary orbital frequency $f_{t}$. Assuming MBHBs evolving in stellar bulges, We take $f_{t}$ from equation (\ref{eq:ft}) assuming the four environment models presented in Section \ref{sec:fttcsingle}. As for $e_t$ we make the simplifying assumption that, regardless of redshift, mass and environment, all MBHBs share the same eccentricity at the transition frequency. We take $e_t=0.01, 0.5, 0.9, 0.99$. 


For each model $h_c(f)$ is computed either via equations (\ref{eq:hc},\ref{eq:dEdf},\ref{eq:dEdt},\ref{eq:fcombined},\ref{eq:ecombined}), i.e., by solving the binary evolution numerically --including the stellar driven phase--  and summing-up all the harmonics, or via equation (\ref{eq:hcanalytic}), i.e., by employing our fitting spectrum for GW driven binaries defined by $f_{t},e_t$.

Results are presented in figures \ref{fig:specpopnum} and \ref{fig:specpopcomp}. Figure \ref{fig:specpopnum} shows the impact of $e_t$ on the spectrum. We notice that in the true spectrum (the dashed lines), changing the population from almost circular to highly eccentric shifts the peak of the spectrum by more then one order of magnitude. As already described, our model does not represent well the low frequency turnover for small $e_t$, however in all cases, the GW signal is well described by equation (\ref{eq:hcanalytic}) in the relevant PTA frequency band ($f>1\,$nHz), and the factor of $\approx 10$ peak shift in the eccentricity range $0.5<e_t<0.99$ is fairly well captured. As anticipated, typical turnover frequencies due to three body scattering are at sub-nHz scales, and flattening (and eventually turnover) in the GW spectrum is observable only if MBHBs have relatively high eccentricities at transition frequency. Figure \ref{fig:specpopcomp} shows the impact of changing the physical parameters describing the efficiency of stellar driven MBHB hardening. Those parameters are fixed to a fiducial value in our model, but can in principle have an impact on the spectrum of the signal. When directly compared to \ref{fig:specpopnum}, the left panel, showing {\it model-NUM}, clarifies that none of those parameters affect the signal to a level comparable to $e_t$. The reason is that essentially all of them cause a change in $\rho_i$, and the $f_t$ dependence on $\rho_i$ is extremely mild ($f_t\propto\rho_i^{3/10}$). For example, shrinking the characteristic galaxy radius $a$ by a factor of two, is equivalent to increasing $\rho_i$ by a factor of eight, which still results in a $<2$ shift of $f_t$. In the right set of panels we see that the same applies to {\it model-AN}. However, there is a striking difference of almost an order of magnitude in the location of the peak. This is because {\it model-NUM} and {\it model-AN} have a very different underlying MBHB mass function. This means that the GWB is dominated by MBHBs with different typical masses, which decouple at different $f_t$. So even if the underlying MBHB dynamics and eccentricity at transition $e_t$ is the same, the resulting peak frequency can be significantly shifted. It is therefore clear that the location of the GWB spectrum turnover is sensitive to both $e_t$ and to the parameters defining the MBHB cosmological mass function, and much less sensitive to the details of the stellar hardening process. This also means, however, that in absence of additional features in the spectrum, the determination of $e_t$ is highly degenerate with the shape of the MBHB mass function.


\subsection{Removal of individual sources}

\begin{figure}
\centering
\includegraphics[width=0.45\textwidth]{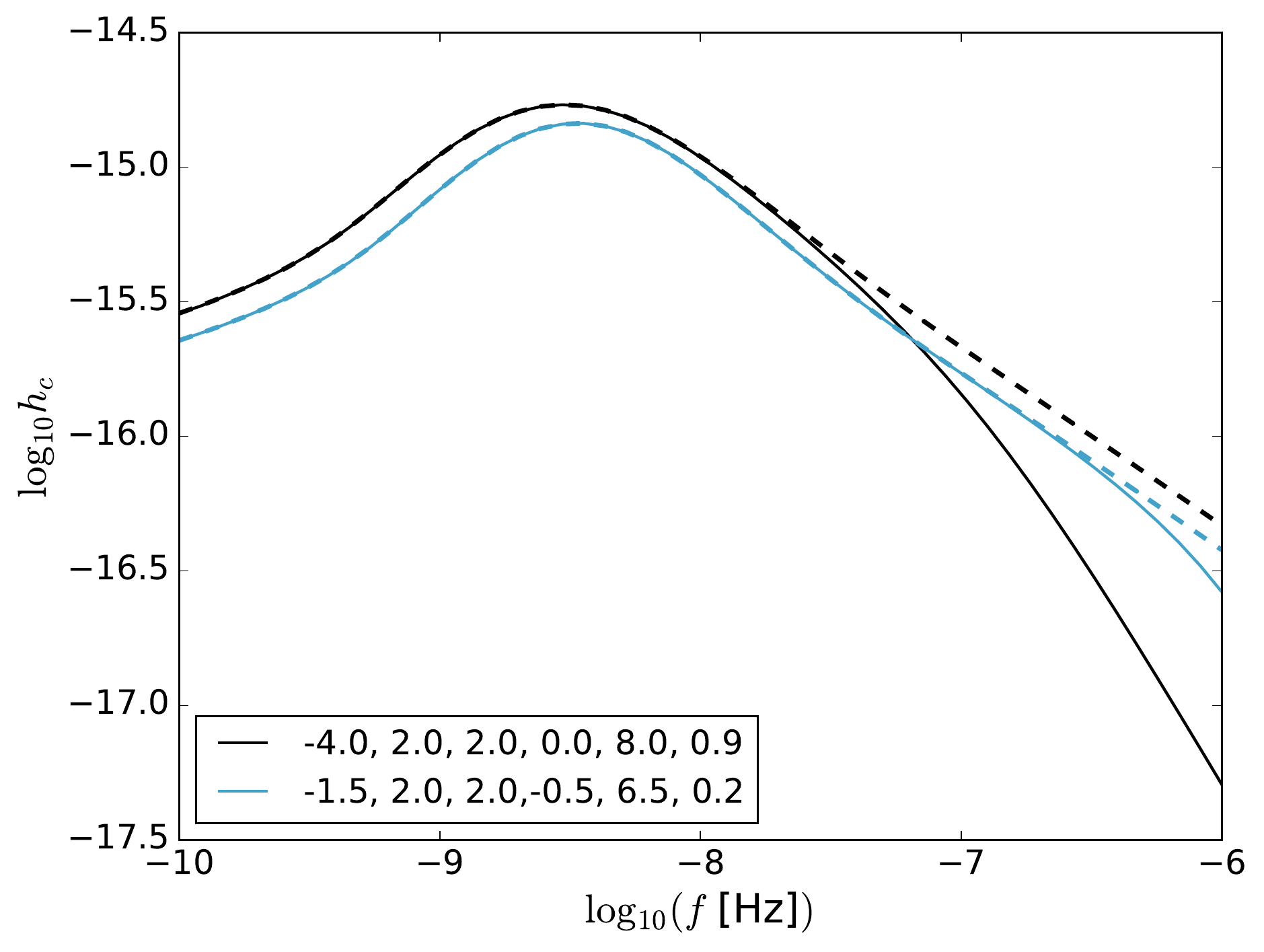}
\caption{Comparison of the spectrum of a population of binaries with different parameters for the {\it model-AN}. Parameters in the plot are specified in the sequence $\{{\rm log}_{10}(\dot{n}_0),\beta,z_*,\alpha,{\rm log}_{10}({\cal M}_*),e_t\}$.
The solid lines represent the spectrum with the drop in upper mass limit in the high frequency regime, the dashed lines represent the spectrum with no mass limit change.}
\label{fig:specmdrop}
\end{figure}

Interestingly, as mentioned in the introduction, another feature appearing in the GW spectrum at high frequencies has been pointed out by SVC08, and depends on the shape of the cosmic MBHB mass function. Let us consider circular binaries. In an actual observation, the GW signal generated by a cosmic population of MBHBs at a given observed frequency bin, is given by the sum of all MBHBs emitting at that frequency. This is related to the cosmic density of merging binaries via standard cosmology transformations:
\begin{equation}
\frac{d^2n}{dzd \log_{10} \mathcal{M}}= \frac{d^3 N}{dz d \log_{10} \mathcal{M} df} \frac{df}{df_r} \frac{df_r}{dt_r} \frac{dt_r}{dz} \frac{dz}{dV_c}
\end{equation}
where \cite{Hogg:1999}
\begin{align}
\frac{dV_c}{dz} & = \frac{4\pi c}{H_0} \frac{D_M^2}{(\Omega_M(1+z)^3 + \Omega_k(1+z)^2 + \Omega_\Lambda)^{1/2}},
\\
D_M & = \frac{c}{H_0} \int_0^z \frac{dz'}{(\Omega_M(1+z')^3 + \Omega_k(1+z')^2 + \Omega_\Lambda)^{1/2}},
\\
\frac{df_r}{dt_r} & = \frac{96}{5} (\pi)^{8/3} \frac{G^{5/3}}{c^5} \mathcal{M}^{5/3} f_r^{11/3},
\\
\frac{df}{df_r} & = \frac{1}{1+z},
\end{align}
and $dt_r/dz$ is given by equation (\ref{eq:dtrdz}). The number of sources emitting in a given observed frequency bin of width $\Delta{f}=1/T$ is therefore given by:
\begin{equation}
  N_{\Delta{f}}=\int_{f-\Delta f/2}^{f+\Delta f/2} \int_0^{\infty} \int_{0}^{\infty} \frac{d^3 N}{df dz d \log_{10} \mathcal{M}}.
  \label{eq:Nf}
\end{equation}
Each chirp mass and redshift bin contribute to $h_c$ in a measure that is proportional to ${\cal M}^{5/6}/(1+z)^{1/6}$ (see, e.g., equation (\ref{eq:hc0})). Therefore, it is possible to rank systems in order of decreasing contribution to the GWB. Because of the very small dependence on redshift -- $1<(1+z)^{1/6}<1.3$ for $0<z<5$ considered in our models -- we simplify the problem by integrating over $z$ and rank systems based on mass only. It is easy to show that $d^2N/{dfd \log_{10} \mathcal{M}}$ is a strong decreasing function of mass, and is in general $\ll 1$ for the most massive systems when $f>10$nHz. This means that the contribution to the GWB coming from those massive sources at that frequency, is in fact given by 'less than one source'. Since the actual GW signal is given by a discrete population of sources, having less than a source in a given frequency bin means that in a typical realization of the Universe that source might or might not be there with a given probability. For the practical purpose of the GWB computation, the contribution from those systems at those frequencies is actually not there, at least not in the form of a stochastic GWB (we defer the reader to SVC08 for a rigorous mathematical treatment of this issue).  

One can therefore assume that in each bin $\Delta{f}$ the most massive sources integrating to $1$ in number do not contribute to the GWB. The value $\bar{M}$ corresponding to this condition is implicitly given by imposing

\begin{equation}
\begin{split}
1 = & \int_{\bar{M}}^{\infty} \int_{f-\Delta f/2}^{f+\Delta f/2} \int_0^{z_{\rm max}} \frac{d^3 N}{df dz d\log_{10} \mathcal{M}}
\\
= & \dot{n}_0 \int_{\bar{M}}^{\infty} \Big(\frac{\mathcal{M}}{10^7 M_\odot}\Big)^{-\alpha} e^{-\mathcal{M}/\mathcal{M}_*} \mathcal{M}^{-5/3} d\log_{10} \mathcal{M}
\\
& \int_0^{\bar{z}} (1+z)^{\beta+1} e^{-z/z_*} \frac{dV_c}{dz} dz \int_{f-\Delta f/2}^{f+\Delta f/2} \frac{dt_r}{df_r} \mathcal{M}^{5/3} df
\end{split}
\label{eq:Mmax}
\end{equation}
Where in the last equation we substituted the analytical merger rate density given by equation (\ref{eq:pop}). Given an observation time $T$, the frequency spectrum is divided in bins $\Delta{f}=1/T$. $h_c(f)$ is therefore calculated at the centroid of each frequency bin by substituting the upper limit $\bar{M}$ defined by equation (\ref{eq:Mmax}) in equation (\ref{eq:hc}). Note that in equation (\ref{eq:Mmax}) mass and frequency integrals are analytic, and only the redshift integral has to be evaluated numerically.

Examples of the GW spectrum obtained including the $\bar{M}$ cut-off are shown in figure \ref{fig:specmdrop} for two different mass functions assuming {\it model-AN}. Note that the spectrum is significantly reduced only at $f>10$nHz. This justifies a posteriori our assumption of  circular GW driven binaries; at such high frequencies even MBHBs that were very eccentric at $f_t$ had become almost circular because of GW backreaction. The figure illustrates that a detection of both spectral features (low frequency turnover and high frequency steepening) might help breaking degeneracies between $e_t$ and MBHB mass function. The two displayed models have very different $e_t$ (0.2 vs 0.9), but also quite different mass functions, so that the GWB turnover occurs around the same frequency. If the signal can be detected up to $f\approx 10^{-7}$Hz, however, differences in the high frequency slope might help pinning down the MBHB mass function and disentangle it from $e_t$. In a companion paper \citep{2017MNRAS.468..404C}, we explore the feasibility of this approach and the implication for astrophysical inference.

\section{Discussion and conclusions}
\label{sec:Conclusions}

In this paper we developed a semi-analytical model that allows the fast computation of the stochastic GWB from a population of eccentric GW driven MBHBs. The spectrum computation does not directly take into account for any coupling of the MBHB with its stellar and gaseous environment and therefore cannot provide a trustworthy description of the GW signal at all frequencies. The coupling enters in the calculation only by setting the characteristic binary population eccentricity $e_t$ at the transition (or decoupling) frequency $f_t$. We showed, however, that in the plausible astrophysical scenario of MBHBs driven by three body scattering of ambient stars, $f_t<1$nHz (for MBHBs with ${\cal M}>10^8\msun$, that dominate the PTA signal), which is a plausible lower limit for future PTA efforts. Therefore environmental coupling only affects the direct computation of the GW signal in a frequency range that is likely inaccessible to current and near future PTAs, justifying our strategy.
Our simple semi-analytic model therefore provides a quick and accurate way to construct the GWB from a population of eccentric MBHBs evolving in stellar environment {\it in the frequency range relevant for PTA} (see figure \ref{fig:specpopnum}).

Compared to the standard $f^{-2/3}$ power-law, the GWB shows two prominent spectral features: i) a low frequency turnover defined by the coupling with the environment and typical eccentricity of the MBHBs at the transition frequency (figure \ref{fig:specpopnum}), and ii) a high frequency steepening due to small number statistics affecting the most massive MBHBs contributing to the GW signal at high frequency (figure \ref{fig:specmdrop}, see SVC08).

We consider stellar driven MBHBs and we employ for the first time in a PTA related investigation realistic density profiles, appropriate for massive elliptical galaxies (which are the typical hosts of PTA sources). For example, both \cite{Sesana:2013CQG} and \cite{2014MNRAS.442...56R} used a simplistic double power-law model matching an isothermal sphere at $r>r_i$ defined by equation (\ref{eq:ricondition}). This model is more centrally concentrated and results in much higher $\rho_i$ and $f_t$ than what found in the present study. We find that in density profiles that are typical for massive ellipticals, MBHBs can coalesce on timescales of few Gyr or less (depending on mass and eccentricity) and the typical transition frequency (from stellar driven to GW driven binaries) is located well below $1\,$nHz. Therefore, an observed turnover in the GWB spectrum in the PTA relevant frequency range is likely to be due to high eccentricities rather than coupling with the environment. In particular, we find that a low frequency bending is likely to be seen for $e_t>0.5$, whereas a proper turnover is characteristic of MBHB populations with $e_t>0.9$. These findings are robust against a variety of plausible host galaxy models; i.e. the properties of the stellar environment affect the location of the bending/turnover of the spectrum only within a factor of two within the cases examined here. This latter point deserves some further consideration. All the physical parameters describing the environment of the MBHB affect the location of $f_t$ through $\rho_i$. Essentially it is the density at the influence radius of the binary (together with the MBHB mass and eccentricity) that determines $f_t$. Although for a range of astrophysically plausible scenarios the typical $\rho_i$ for a given MBHB is found to vary within a factor of ten, it might be worth considering the possibility of more extreme scenarios. This can be easily incorporated in our treatment as a free multiplicative parameter to $\rho_i$, and we plan to expand our model in this direction in future investigations. 

This has a number of interesting consequences in terms of astrophysical inference from PTA observations. Firstly, for $e_t<0.5$ no low frequency signature in the GWB spectrum is likely to be seen, making it impossible to distinguish circular from mildly eccentric MBHBs {\it on the basis of the GWB spectral shape only}. Secondly, because of the $(\rho_i/\sigma)^{3/10}$ dependence of equation (\ref{eq:ft}) it will be difficult to place strong constraints on the stellar environment of MBHBs via PTA observations. Lastly, a turnover in the PTA band would be indicative of an highly eccentric ($e_t>0.9$) MBHB population. The turnover frequency depends on both $e_t$ and the MBHB mass function (through the ${\cal M}$ dependence of  $f_t$), therefore the detection of a low frequency turnover alone might not place strong constraints on the typical MBHB eccentricity. The high frequency steepening, on the other hand, generally occur at $f>10\,$nHz, where MBHBs have mostly circularized. Therefore it depends exclusively on the MBHB mass function. A measurement of such steepening can therefore constrain the MBHB mass function and break the mass function eccentricity degeneracy affecting the location of the low frequency turnover.

Looking at the prospects of performing astrophysical inference from PTA data, our model has several advantages. First, it directly connects the relevant astrophysical parameters of the MBHB population to the shape of the GWB. As mentioned above, in this first paper, we keep the MBHB mass function and eccentricity at decoupling as free parameters, arguing that other factors affecting the dynamics likely have a minor impact on the signal. Those can, however, be incorporated in our scheme as additional free parameters, if needed. This will eventually allow to perform astrophysical inference from PTA measurements exploiting a model that self consistently includes all the relevant physics defining the MBHB population. This improves upon the 'proof of principle' type of analysis performed in \cite{ArzoumanianEtAl_NANOGRAV9yrData:2016}, where limits on different model ingredients were placed by adding them individually to the model. For example, by assuming a standard $f^{-2/3}$ power-law, limits were placed on the MBH-host relation. Then a prior on the amplitude was assumed and an ad-hoc broken power law was constructed to put constraints on environmental coupling. Finally, the latter was put aside and eccentricity was added to the model to be constrained separately. Although this is a useful exercise, eventually all ingredients have to be considered at the same time, to be meaningfully constrained, and our modelling takes a step in this direction. Second, the model is mostly analytical, involving only few numerical integrals. The most computationally expensive operations, namely the integration of the MBHB orbital evolution and the summation over all the harmonics of the GW signal, are captured by the simple fitting formula given in equation (\ref{eq:hcfit}), together with its simple scaling properties. Therefore, for a given set of parameters $\{{\rm log}_{10}(\dot{n}_0),\beta,z_*,\alpha,{\rm log}_{10}({\cal M}_*),e_t\}$ the GWB can be numerically computed within few ms. This makes the model suitable for large parameter space exploration via parallel Markov Chain Monte Carlo or Nested Sampling searches. In a companion paper \citep{2017MNRAS.468..404C} we explore this possibility and demonstrate which MBHB parameters and with which accuracy, one can constrain from PTA observations.

Before closing we stress again that we consider the {\it shape of the GWB only}. Additional information about the MBHB population will be enclosed in the statistical nature of this background (whether, for example, it is non-Gaussian, anisotropic, non-stationary) and in the properties of individually resolvable sources. A comprehensive account of astrophysical inference from PTA observations will necessarily have to take into account for all this combined information, and our current investigation is only the first step in this direction.

\section*{acknowledgements}
We acknowledge the support of our colleagues in the European Pulsar Timing Array. A.S. is supported by a University Research Fellow of the Royal Society. 



\bibliographystyle{mnras}
\bibliography{bibliography}

\label{lastpage}

\end{document}